\documentclass[english]{article}
\usepackage[LGR,LGR,LGR,LGR,LGR,T1]{fontenc}
\usepackage[latin9]{inputenc}
\usepackage{geometry}
\usepackage{color}
\usepackage{array}
\usepackage{float}
\usepackage{amsmath}
\usepackage{amssymb}
\usepackage{graphicx}
\usepackage{epstopdf}
\usepackage{epsfig}

\makeatletter

\providecommand{\tabularnewline}{\\}

\usepackage{babel}

\makeatother

\begin{document}

\title{\textbf{Hierarchical Joint Remote State Preparation in Noisy
Environment}}

\author{Chitra Shukla$^{a,}$\thanks{email: shukla.chitra@i.mbox.nagoya-u.ac.jp },
Kishore Thapliyal$^{b,}$\thanks{email: tkishore36@yahoo.com}, Anirban
Pathak$^{b,}$\thanks{email: anirban.pathak@jiit.ac.in}}

\maketitle
\begin{center}
$^{a}$Graduate School of Information Science, Nagoya University,
Furo-cho 1, Chikusa-ku, Nagoya, 464-8601, Japan
\par\end{center}

\begin{center}
$^{b}$Jaypee Institute of Information Technology, A-10, Sector-62,
Noida, UP-201307, India
\par\end{center}

\begin{abstract}
A novel scheme for quantum communication having substantial applications in practical life is designed and analyzed. Specifically, we have proposed a hierarchical counterpart of the joint remote state
preparation (JRSP) protocol, where two senders can jointly and remotely
prepare a quantum state. One sender has the information regarding
amplitude, while the other one has the phase information of a quantum state to be jointly prepared at the receiver's port.
However, there exists a hierarchy among the receivers, as far as powers
to reconstruct the quantum state is concerned. A 5-qubit cluster state has been used here
to perform the task. Further, it is established that the proposed
scheme for hierarchical JRSP (HJRSP) is of enormous practical importance
in critical situations involving defense and other sectors, where
it is essential to ensure that an important decision/order that can
severely affect a society or an organization is not taken by a single
person, and once the order is issued all the receivers don't possess
an equal right to implement it. Further, the effect of different noise
models (e.g., amplitude damping (AD), phase damping (PD), collective
noise and Pauli noise models) on the HJRSP protocol proposed here
is investigated. It is found that in AD and PD noise models a higher
power agent can reconstruct the quantum state to be remotely prepared
with higher fidelity than that done by the lower power agent(s). In
contrast, the opposite may happen in the presence of collective noise
models. We have also proposed a scheme for probabilistic HJRSP using
a non-maximally entangled 5-qubit cluster state. 
\end{abstract}
\textbf{Keywords:} Joint remote state preparation, hierarchical quantum
communication, amplitude damping noise, phase damping noise, collective
noise, Pauli noise, quantum communication.

\section{Introduction}

There are two major facets of quantum information science: quantum
computing \cite{Grover's algorithm,Shor-algo,nielsen} and quantum
communication \cite{nielsen,BB84,Bennet-Brassard-Teleportation,my-book}.
In the last three decades, both of these have attracted major attraction
of the scientific community for their ability to perform computational
or communication tasks beyond the capacity of their classical counterparts.
Specifically, a quantum algorithm may perform a computational task
much faster than its classical counterpart. For example, Grover's
algorithm \cite{Grover's algorithm} can search unsorted database
quadratically faster than its best known classical counterpart, and
Shor's algorithm \cite{Shor-algo} can factorize large numbers in
a speed not attainable in a classical computer. Similarly, in the
domain of quantum communication, protocols have been proposed for teleportation
\cite{Bennet-Brassard-Teleportation}, which does not have any classical
analogue, and unconditional security of quantum key distribution \cite{BB84}
has been established \cite{BB84-secPr}. Unconditional security is a desirable
feat not achievable in classical world. The nonclassical nature of
quantum communication schemes drew considerable attention of the scientific
community. Especially, teleportation (a communication scheme, in which
the state to be transmitted from the sender to the receiver never
exists in the channel joining the receiver and the sender) drew major
attention because of its magical characteristics. Initially, a teleportation
scheme was proposed by Bennett et al. in 1993 \cite{Bennet-Brassard-Teleportation}.
In the original teleportation scheme, the sender Alice used to transmit/teleport
an unknown single-qubit quantum state to the receiver Bob using a
shared entangled state and two bits of classical communication. Subsequently,
several variants of teleportation have been proposed. For example,
schemes were proposed for quantum information splitting (QIS) or controlled
teleportation (CT) \cite{Ct,A.Pathak}, quantum secret sharing (QSS)
\cite{Hillery}, hierarchical quantum information splitting (HQIS)
\cite{hierarchical,Shukla}, remote state preparation (RSP) \cite{Pati},
etc. (see Ref. \cite{my-book} for a review). All these schemes can
be viewed as variants of teleportation.

Recently, a few hierarchical versions of already existing aspects
of quantum communication (variants of teleportation) have been proposed.
Specifically, hierarchical quantum information splitting (HQIS) \cite{hierarchical,Shukla,Wang1,Wang2},
hierarchical quantum secret sharing (HQSS) \cite{Shukla}, hierarchical
dynamic quantum secret sharing (HDQSS) \cite{HDQSS}, etc., have been
proposed in the recent past. It is also shown that these schemes have
enormous practical importance (for a detailed discussion on the interesting
applications of these schemes see Sec. 1 of Refs. \cite{Shukla,HDQSS}).
In these protocols, there is a hierarchy among the powers of receivers
(agents) to reconstruct a quantum state sent by the sender, i.e.,
the agents are graded in accordance to their power for the reconstruction
of an unknown quantum state. Specifically, in HQIS the receivers can
reconstruct the teleported quantum state with the help of other receivers
(as in QIS \cite{Ct,A.Pathak}), where the power of a particular receiver
is decided by the minimum number of receivers required to cooperate
with him to enable him to reconstruct the state \cite{Shukla}. In
probabilistic HQIS, the same task is performed probabilistically.
HQSS scheme can be viewed as a direct application of HQIS where the
sender wishes to send an information in pieces to all the receivers
who can reconstruct it with the help of either some or all other agents.
Later the scheme was extended to propose a HDQSS scheme, and in HDQSS
scheme an additional feature to add and drop an agent was included
\cite{HDQSS}, and that made HDQSS most practical hierarchical scheme
proposed until now. This is so because, in a practical situation,
an agent may resign from a company or take leave, and the company may decide
to recruit a new agent as his/her replacement. Further, if the sale of the company gets increased it may recruit a new agent. However, these recently introduced
schemes of hierarchical quantum communication lags a particularly important
feature. In all the existing schemes, there is only one sender, but
in many practical purposes (as illustrated with an example in Sec.
\ref{sec:Practical-Applications}) we need more than one sender for
the secrecy of the initial message/state to be transmitted in a\textcolor{magenta}{{}
}hierarchical manner. This paper aims to address this particular issue
and design a new type of scheme for hierarchical quantum communication,
which we would refer to as hierarchical joint remote state preparation (HJRSP)
scheme in analogy with the well known joint remote state preparation (JRSP)
schemes \cite{Ba-An-RSP,RSP-arbitrary-bipartite-usingGHZ-type,JRSP-W,JRSP-W-state}. 

Here, it would be apt to note that there exists a variant of teleportation
known as RSP, where a known quantum state is remotely prepared at
the receiver's end. Thus, RSP may be viewed as teleportation of a
known quantum state. The first RSP scheme was proposed by Pati \cite{Pati}
using Bell states, in 2000. This scheme required 1 bit of classical
communication and a shared entangled state (1 ebit). As the standard
teleportation scheme requires 2 cbit and 1 ebit, Pati's scheme of
RSP was able to teleport a known quantum state with reduced resources
compared to the unknown qubit case. However, this scheme was probabilistic
in nature, as the success probability was not unity. Later a deterministic
counterpart of the RSP scheme was proposed by An et al. \cite{BaAnRSP},
where the required resources become equal to that of teleportation
\cite{Bennet-Brassard-Teleportation}. Subsequently, a large amount
of work has been carried out on RSP. In these works RSP has been implemented in probabilistic, deterministic,
controlled and controlled bidirectional manner using different quantum
states, such as $n$-level, 4-qubit $GHZ$ state, multi-qubit $GHZ$
state, 4-qubit cluster-type state, arbitrary two qubit state, and
$W$ state \cite{Ba-An-RSP,RSP-arbitrary-bipartite-usingGHZ-type,JRSP-W,JRSP-W-state,CBRSP,New0_RSP_2013,RSP-1-GHZ,RSP-multiparitite-ghz,RSP-4-qubit-cluster-type,RSP-cluster-type}.
Interestingly, a few experimental realizations of some of the RSP
schemes have also been reported in the past \cite{EXp-first experimental RSP,Exp-RSP-optical1,EXP-RSP-optical2,EXP-RSP-Experiment-mixed-state,Expt. multilocation-RSP,Expt-RSP-NMR}. 

In 2008, the RSP scheme was modified to a three-party scheme of the
joint RSP (JRSP) \cite{Ba-An-RSP}. This one is a unique quantum communication
scheme, where two senders jointly prepare a known quantum state at
the remote port. The state to be prepared remotely is neither completely
known to sender 1 nor to sender 2, but they jointly know the state
to be teleported. It is worth stressing that this scheme has no analogue
in teleportation, as an unknown quantum state cannot be teleported
in this way by more than one sender. Since then several schemes for
JRSP have been proposed \cite{RSP-arbitrary-bipartite-usingGHZ-type,JRSP-W,JRSP-W-state}.
In fact, the JRSP \cite{RSP-with-noise} scheme was investigated under
the amplitude damping (AD) and phase damping (PD) noise models in
the recent past. Interestingly, practical applications of the hierarchical
quantum communication schemes, described in the context of HQIS \cite{Shukla}
and HDQSS \cite{HDQSS}, motivated us to investigate the possibility
of designing a hierarchical version of JRSP (HJRSP) scheme with the
hope of using it in some of the practical purposes of daily life.
Further, recent investigations on RSP schemes \cite{CBRSP,RSP-with-noise}
under the noisy environment, such as AD and PD channels, motivated
us to simulate a similar study for the proposed HJRSP protocol.\textcolor{red}{{} }

\begin{figure}
\begin{centering}
\includegraphics[scale=0.55,angle=270]{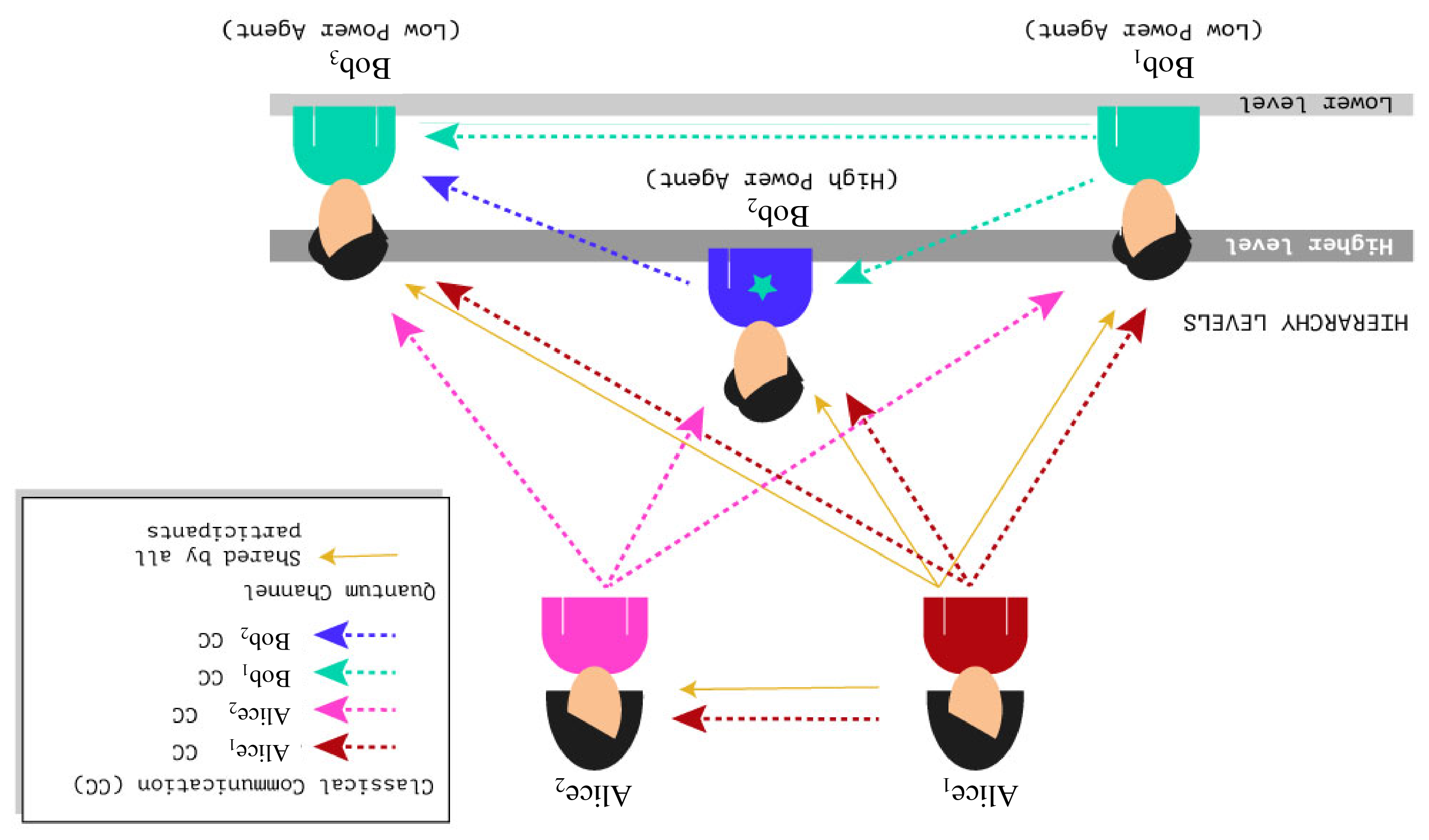}
\par\end{centering}

\protect\caption{\label{fig:HJRSP-scheme}HJRSP scheme illustrated through a schematic diagram.
The quantum and classical communications involved between all the parties
are shown using the smooth and dashed lines, respectively. It is shown that $\rm{Bob_{2}}$ can reconstruct the quantum state with the help of only one of the lower power agents (in the diagram it is shown as $\rm{Bob_{1}}$). In contrast, $\rm{Bob_{3}}$ requires help of both $\rm{Bob_{1}}$ and $\rm{Bob_{2}}$. Here, we have not shown the case when $\rm{Bob_{1}}$ reconstructs the quantum state as it is similar to the case when $\rm{Bob_{3}}$ does so}. 
\end{figure}

The HQIS scheme \cite{Shukla} proposed in the recent past involves
teleportation of an unknown quantum state among three receivers
hierarchically. Further, we know that in a JRSP scheme,\textcolor{magenta}{{}
}there are two senders having the information of amplitude and relative
phase, respectively. Hence, the proposed HJRSP scheme should have at least
five parties (2 senders and three receivers to ensure joint preparation and hierarchical reconstruction), and therefore, we need at least a 5-qubit quantum state
for the implementation of the HJRSP protocol. This is why, we have used a
5-qubit cluster state of the form

\begin{equation}
|C\rangle=\frac{1}{2}\left[|00000\text{\ensuremath{\rangle+|00111\rangle+|11010\rangle+|11101\rangle}}\right]_{S_{1}S_{2}R_{1}R_{2}R_{3}}\label{eq:state}
\end{equation}
with the qubits $S_{1},\,S_{2}$ belonging to the senders ${\rm Alice}{}_{1}$
and ${\rm Alice}{}_{2}$; and $R_{1},\,R_{2}$ and $R_{3}$ corresponding
to the qubits of the three receivers, who are referred to as ${\rm Bob_{1},}\,{\rm Bob_{2},\,and\,Bob_{3}}$,
respectively. The preparation of the cluster state (\ref{eq:state})
by ${\rm Alice}{}_{1}$ and distribution of qubits to all other parties
is illustrated through the schematic diagram shown in Fig. \ref{fig:HJRSP-scheme}.
More detail of the scheme follows in the forthcoming
section.

In the following sections, the paper is organized as follows. In Sec.
\ref{sec:Hierarchical-Joint-Remote}, we propose a protocol of deterministic
HJRSP using 5-qubit cluster state of the form (\ref{eq:state}). This
is followed by a protocol of probabilistic HJRSP using a non-maximally
entangled counterpart of the 5-qubit cluster state (\ref{eq:state}).
Subsequently, we have discussed some practical applications of the
proposed HJRSP schemes in Sec. \ref{sec:Practical-Applications}.
Further, in Sec. \ref{sec:Effect-of-set}, we study the effect of
a set of noise models on the proposed deterministic HJRSP scheme.
Finally, we conclude the paper in Sec. \ref{sec:Conclusions}.

\section{Hierarchical joint remote state Preparation with 5-qubit cluster
state\label{sec:Hierarchical-Joint-Remote}}

In a HJRSP scheme, we have two senders (${\rm Alice_{1}}$ and ${\rm Alice_{2}}$),
such that ${\rm Alice_{1}}$ knows the information about the amplitude,
and ${\rm Alice_{2}}$ has the phase information related to the qubit
$|\xi\rangle=a|0\rangle+be^{i\phi}|1\rangle$ : $a^{2}+b^{2}=1$.
It implies that the senders jointly know the quantum state to be prepared
at the remote location (i.e., at the receivers' end), whereas it is unknown
to the receivers. Further, we require at least three receivers to
design a hierarchical scheme for joint remote state preparation. Here,
the three receivers are referred to as ${\rm Bob_{1}}$, ${\rm Bob_{2}}$
and ${\rm Bob_{3}}$, respectively. The hierarchy among the receivers
will be established in what follows. To accomplish the task, we have
chosen a 5-qubit cluster state of the form (\ref{eq:state}) where
each of the participants possesses one qubit of the 5-qubit cluster
state. Subsequently, a probabilistic version of HJRSP scheme is introduced,
and it's shown that the probabilistic HJRSP scheme appears when we
use a non-maximally entangled state of the form (\ref{eq:state})
as a quantum channel.

\subsection{Deterministic HJRSP}

We may now describe a protocol of deterministic HJRSP using a quantum
state of the form (\ref{eq:state}) in following steps:
\begin{description}
\item [{Step~1}] ${\rm Alice}_{1}$ prepares the 5-qubit cluster state (\ref{eq:state}).
She keeps $S_{1}$ ($1^{{\rm st}}$) qubit with herself and sends
$S_{2}$ ($2^{{\rm nd}})$ qubit to ${\rm Alice_{2}}$ and $R_{1},\,R_{2},\,R_{3}$
qubits ($3^{{\rm rd}},\,4^{{\rm th}}\,{\rm and}\,5^{{\rm th}}$ qubits)
to the three receivers ${\rm Bob_{1}}$, ${\rm Bob_{2}}$, ${\rm Bob_{3}}$,
respectively.
\item [{Step~2}] ${\rm Alice}_{1}$ measures her qubit in $\{u_{0},u_{1}\}$
basis, where $|u_{0}\rangle=a|0\rangle+b|1\rangle$ and $|u_{1}\rangle=b|0\rangle-a|1\rangle$,
and announces the measurement outcome. \\
As the cluster state (\ref{eq:state}) can be expressed as 
\begin{equation}
\begin{array}{lcl}
|C\rangle & = & \frac{1}{2}\left[|u_{0}\rangle_{S_{1}}\{a(|0000\rangle+|0111\rangle)+b(|1010\rangle+|1101\rangle)\}_{S_{2}R_{1}R_{2}R_{3}}\right.\\
 & + & \left.|u_{1}\rangle_{S_{1}}\{b(|0000\rangle+|0111\rangle)-a(|1010\rangle+|1101\rangle)\}_{S_{2}R_{1}R_{2}R_{3}}\right],
\end{array}\label{eq:det-step2}
\end{equation}
the measurement of ${\rm Alice}_{1}$ would reduce $|C\rangle$ to
$\text{|}\varPsi_{0}\rangle\left(\text{|}\varPsi_{1}\rangle\right)$
if her measurement outcome is $|u_{0}\rangle(|u_{1}\rangle)$, where
\begin{equation}
\begin{array}{lcl}
\text{|}\varPsi_{0}\rangle & = & \frac{1}{\sqrt{2}}\{a(|0000\rangle+|0111\rangle)+b(|1010\rangle+|1101\rangle)\}_{S_{2}R_{1}R_{2}R_{3}},\\
|\varPsi_{1}\rangle & = & \frac{1}{\sqrt{2}}\left\{ b(|0000\rangle+|0111\rangle)-a(|1010\rangle+|1101\rangle)\right\} {}_{S_{2}R_{1}R_{2}R_{3}}.
\end{array}\label{eq:psi0-1}
\end{equation}

\item [{Case~1:}] \textbf{In Step 2, ${\rm Alice}_{1}$'s measurement yields $|u_{0}\rangle$ }
\item [{Step~3}] If ${\rm Alice}_{1}$'s measurement yields $|u_{0}\rangle$
then ${\rm Alice}_{2}$ applies a single qubit phase gate $P(2\phi)=\left(\begin{array}{lc}
1 & 0\\
0 & e^{2i\phi}
\end{array}\right)$ on her qubit and that transforms $\text{|}\varPsi_{0}\rangle$ to
\begin{equation}
\begin{array}{lcl}
\text{|}\varPsi_{0}^{\prime}\rangle & = & \left(P(2\phi)\otimes I_{2}^{\otimes3}\right)\text{|}\varPsi_{0}\rangle=\frac{1}{\sqrt{2}}\{a(|0000\rangle+|0111\rangle)+be^{2i\phi}(|1010\rangle+|1101\rangle)\}_{S_{2}R_{1}R_{2}R_{3}}\\
 & = & \frac{1}{2}\left[|v_{0}\rangle_{S_{2}}\{a(|000\rangle+|111\rangle)+be^{i\phi}(|010\rangle+|101\rangle)\}_{R_{1}R_{2}R_{3}}\right.\\
 & + & \left.e^{i\phi}|v_{1}\rangle_{S_{2}}\{a(|000\rangle+|111\rangle)-be^{i\phi}(|010\rangle+|101\rangle)\}_{R_{1}R_{2}R_{3}}\right]\\
 & = & \frac{1}{\sqrt{2}}\left(|v_{0}\rangle_{S_{2}}|\phi_{0}\rangle_{R_{1}R_{2}R_{3}}+|v_{1}\rangle_{S_{2}}|\phi_{1}\rangle_{R_{1}R_{2}R_{3}}\right),
\end{array}\label{eq:det-step3}
\end{equation}
where $|v_{0}\rangle=\frac{|0\rangle+e^{i\phi}|1\rangle}{\sqrt{2}}$,
$|v_{1}\rangle=\frac{e^{^{-i\phi}}|0\rangle-|1\rangle}{\sqrt{2}}$,
$|\phi_{0}\rangle=\frac{1}{\sqrt{2}}\left\{ a(|000\rangle+|111\rangle)+be^{i\phi}(|010\rangle+|101\rangle)\right\} _{R_{1}R_{2}R_{3}}$,
and $|\phi_{1}\rangle=\frac{1}{\sqrt{2}}\left\{ a(|000\rangle+|111\rangle)-be^{i\phi}(|010\rangle+|101\rangle)\right\} _{R_{1}R_{2}R_{3}}$. 
\item [{Step~4}] ${\rm Alice}_{2}$ measures her qubit $S_{2}$ in $\{v_{0},v_{1}\}$
basis and announces the outcome obtained by her. \\
From Eq. (\ref{eq:det-step3}) we can easily observe that if ${\rm Alice}_{2}$
obtains $v_{0}\,(v_{1})$ then the composite state shared by the receivers
reduces to $|\phi_{0}\rangle\left(|\phi_{1}\rangle\right)$.
\item [{Case~1.1:}] \textbf{The receivers decide that ${\rm Bob_{2}}$
will reconstruct the state}
\item [{Step~5}] If the agents decide that \textbf{${\rm Bob_{2}}$ }will
reconstruct the unknown quantum state $|\xi\rangle$, then $|\phi_{0}\rangle$
and $|\phi_{1}\rangle$ can be decomposed as
\end{description}
\begin{equation}
\begin{array}{lcl}
|\phi_{0}\rangle & = & \frac{1}{\sqrt{2}}\left\{ |00\rangle_{R_{1}R_{3}}(a|0\rangle+be^{i\phi}|1\rangle)_{R_{2}}+|11\rangle_{R_{1}R_{3}}(a|1\rangle+be^{i\phi}|0\rangle)_{R_{2}}\right\} ,\\
|\phi_{1}\rangle & = & \frac{1}{\sqrt{2}}\left\{ |00\rangle_{R_{1}R_{3}}(a|0\rangle-be^{i\phi}|1\rangle)_{R_{2}}+|11\rangle_{R_{1}R_{3}}(a|1\rangle-be^{i\phi}|0\rangle)_{R_{2}}\right\} .
\end{array}\label{eq:det-step4}
\end{equation}
From Eq. (\ref{eq:det-step4}), it is clear that \textbf{${\rm Bob_{2}}$
}can reconstruct the quantum state $|\xi\rangle$ with the collaboration
of either \textbf{${\rm Bob_{1}}$ }or \textbf{${\rm Bob_{3}}$ }by
applying Pauli operations $I,\,X,\,Z$ and $iY$, respectively, as
shown in Table \ref{tab:Relation-between-the}. It is important to
note that \textbf{${\rm Bob_{2}}$ }needs the collaboration of either
the receiver \textbf{${\rm Bob_{1}}$ }or \textbf{${\rm Bob_{3}}$}.
He does not require collaboration of both as the measurement performed
by the receivers ${\rm Bob_{1}}$ and ${\rm Bob_{3}}$ in computational
basis (i.e., $\{|0\rangle,|1\rangle\}$) always yield the same outcome
(cf. Column 3 of Table \ref{tab:Relation-between-the}). Thus, the
communication of the measurement outcome from anyone of the receivers
\textbf{${\rm Bob_{1}}$ }or \textbf{${\rm Bob_{3}}$ }is sufficient
for the receiver \textbf{${\rm Bob_{2}}$}. Consequently, the receiver
\textbf{${\rm Bob_{2}}$} is the higher power agent in the proposed
HJRSP scheme.

\begin{table}[H]
\begin{centering}
\begin{tabular}{|>{\centering}p{3cm}|>{\centering}p{3cm}|>{\centering}p{3cm}|>{\centering}p{3cm}|}
\hline 
${\rm Alice}_{1}$'s measurement outcome in $\{u_{0},u_{1}\}$ basis  & ${\rm Alice_{2}}$'s measurement outcome in $\{v_{0},v_{1}\}$ basis  & ${\rm Bob_{i}}$'s measurement outcome (in
$\{|0\rangle,|1\rangle\}$ basis), where $i\in\left\{1,3\right\}$  & Pauli operation to be applied by ${\rm Bob_{2}}$\tabularnewline
\hline 
 & $|v_{0}\rangle$  & $|0\rangle$  & $I$\tabularnewline
\cline{3-4} 
$|u_{0}\rangle$  &  & $|1\rangle$  & $X$\tabularnewline
\cline{2-4} 
 & $|v_{1}\rangle$  & $|0\rangle$  & $Z$\tabularnewline
\cline{3-4} 
 &  & $|1\rangle$  & $iY$\tabularnewline
\hline 
\end{tabular}
\par\end{centering}

\protect\caption{\label{tab:Relation-between-the}The relation between the measurement
outcomes of the senders and receivers with the unitary operations
to be applied by the receiver ${\rm Bob_{2}}$ to recover the quantum
state jointly sent by the senders ${\rm Alice}_{1}$ and ${\rm Alice_{2}}$.
${\rm Bob_{2}}$ needs the collaboration of either of
the agents ${\rm Bob_{1}}$ or ${\rm Bob_{3}}$.}
\end{table}

\begin{description}
\item [{Case~1.2:}] \textbf{The receivers decide that ${\rm Bob_{3}}$
will reconstruct the state}
\end{description}
\textbf{Step~$\mathbf{5^{\prime}}$ }In case the agents wish ${\rm Bob_{3}}$
to reconstruct $|\xi\rangle$, then $|\phi_{0}\rangle$ and $|\phi_{1}\rangle$
can be decomposed as

\begin{equation}
\begin{array}{lcl}
|\phi_{0}\rangle & = & \frac{1}{2}\left[\left(|+\rangle|0\rangle\right)_{R_{1}R_{2}}\left(a|0\rangle+be^{i\phi}|1\rangle\right)_{R_{3}}+\left(|-\rangle|0\rangle\right)_{R_{1}R_{2}}\left(a|0\rangle-be^{i\phi}|1\rangle\right)_{R_{3}}\right.\\
 & + & \left.\left(|+\rangle|1\rangle\right)_{R_{1}R_{2}}\left(a|1\rangle+be^{i\phi}|0\rangle\right)_{R_{3}}-\left(|-\rangle|1\rangle\right)_{R_{1}R_{2}}\left(a|1\rangle-be^{i\phi}|0\rangle\right)_{R_{3}}\right],\\
|\phi_{1}\rangle & = & \frac{1}{2}\left[\left(|+\rangle|0\rangle\right)_{R_{1}R_{2}}\left(a|0\rangle-be^{i\phi}|1\rangle\right)_{R_{3}}+\left(|-\rangle|0\rangle\right)_{R_{1}R_{2}}\left(a|0\rangle+be^{i\phi}|1\rangle\right)_{R_{3}}\right.\\
 & + & \left.\left(|+\rangle|1\rangle\right)_{R_{1}R_{2}}\left(a|1\rangle-be^{i\phi}|0\rangle\right)_{R_{3}}-\left(|-\rangle|1\rangle\right)_{R_{1}R_{2}}\left(a|1\rangle+be^{i\phi}|0\rangle\right)_{R_{3}}\right].
\end{array}\label{eq:step4-2v0}
\end{equation}

From Eq. (\ref{eq:step4-2v0}), it is clear that ${\rm Bob_{3}}$
can reconstruct the quantum state $|\xi\rangle$ by applying the Pauli
operations as shown in Table \ref{tab:Relation-between-the-1} iff
both the receivers ${\rm Bob_{1}}$ and ${\rm Bob_{2}}$ cooperate
simultaneously by measuring their qubits (in $\{|+\rangle,\,|-\rangle\}$
basis and computational basis, respectively) and sharing the measurement
outcomes. This fact can be established mathematically by tracing over
$R_{1}$ and $R_{2}$ qubits, which gives a completely mixed state
for the receiver ${\rm Bob_{3}}$. The measurement outcomes of the
receivers ${\rm Bob_{1}}$ and ${\rm Bob_{2}}$ are summarized in
Column 3 of Table \ref{tab:Relation-between-the-1} with the corresponding
Pauli operations ${\rm Bob_{3}}$ has to apply in the next column.
Here, it is important to note that ${\rm Bob_{3}}$ needs the collaboration
of both ${\rm Bob_{1}}$ and ${\rm Bob_{2}}$, whereas in Case 1.1,
we have seen that ${\rm Bob_{2}}$ can reconstruct the quantum state
with the help of either ${\rm Bob_{1}}$ or ${\rm Bob_{3}}.$ Thus,
${\rm Bob_{2}}$ is a higher power agent. A similar analysis would
reveal that ${\rm Bob_{1}}$ is also a lower power agent as to reconstruct
the unknown quantum state $|\xi\rangle$ sent by the senders he will
also require the help of the remaining two Bobs. 

\begin{table}[H]
\begin{centering}
\begin{tabular}{|>{\centering}p{3cm}|>{\centering}p{3cm}|>{\centering}p{3cm}|>{\centering}p{3cm}|}
\hline 
${\rm Alice}_{1}$'s measurement outcome in $\{u_{0},u_{1}\}$ basis  & ${\rm Alice_{2}}$'s measurement outcome in $\{v_{0},v_{1}\}$ basis  & ${\rm Bob_{1}}$'s and ${\rm Bob_{2}}$'s joint measurement outcome
(in $\{|+\rangle,|-\rangle\}$ and $\{|0\rangle,|1\rangle\}$ bases,
respectively) & Pauli operation to be applied by ${\rm Bob_{3}}$\tabularnewline
\hline 
 &  & $|+\rangle|0\rangle$  & $I$\tabularnewline
\cline{3-4} 
 & $|v_{0}\rangle$  & $|-\rangle|0\rangle$  & $Z$\tabularnewline
\cline{3-4} 
 &  & $|+\rangle|1\rangle$  & $X$\tabularnewline
\cline{3-4} 
$|u_{0}\rangle$  &  & $|-\rangle|1\rangle$  & $iY$\tabularnewline
\cline{2-4} 
 &  & $|+\rangle|0\rangle$  & $Z$\tabularnewline
\cline{3-4} 
 & $|v_{1}\rangle$  & $|-\rangle|0\rangle$  & $I$\tabularnewline
\cline{3-4} 
 &  & $|+\rangle|1\rangle$  & $iY$\tabularnewline
\cline{3-4} 
 &  & $|-\rangle|1\rangle$  & $X$\tabularnewline
\hline 
\end{tabular}
\par\end{centering}

\protect\caption{\label{tab:Relation-between-the-1}The measurement outcomes of all
the senders and receivers, and corresponding Pauli operations ${\rm Bob_{3}}$
has to apply to recover the quantum state. The receiver ${\rm Bob_{3}}$
needs the joint collaboration of all other agents.}
\end{table}

\begin{description}
\item [{Case~2:}] \textbf{In Step2, ${\rm Alice}_{1}$'s measurement yields
$|u_{1}\rangle$ }
\end{description}
If ${\rm Alice_{1}}$'s measurement yields $|u_{1}\rangle$, then
\textbf{Step3 }and the following steps described above will be modified
to \textbf{Step3-1} and so on as described in the following steps:
\begin{description}
\item [{Step~3-1}] If ${\rm Alice_{1}}$'s measurement outcome is $|u_{1}\rangle$
then ${\rm Alice_{2}}$ need not apply the phase gate $P(2\phi)$,
and the state can be written as
\begin{equation}
\begin{array}{lcl}
|\varPsi_{1}\rangle & = & \frac{1}{\sqrt{2}}\left[b(|0000\rangle+|0111\rangle)-a(|1010\rangle+|1101\rangle)\right]_{S_{2}R_{1}R_{2}R_{3}}\\
 & = & \frac{1}{2}\left[|v_{0}\rangle_{S_{2}}\{b(|000\rangle+|111\rangle)-ae^{-i\phi}(|010\rangle+|101\rangle)\}_{R_{1}R_{2}R_{3}}\right.\\
 & + & \left.|v_{1}\rangle_{S_{2}}\{be^{i\phi}(|000\rangle+|111\rangle)+a(|010\rangle+|101\rangle)\}_{R_{1}R_{2}R_{3}}\right]\\
 & = & \frac{1}{\sqrt{2}}\left[|v_{0}\rangle_{S_{2}}|\phi_{0}^{\prime}\rangle_{R_{1}R_{2}R_{3}}+|v_{1}\rangle_{S_{2}}|\phi_{1}^{\prime}\rangle_{R_{1}R_{2}R_{3}}\right].
\end{array}\label{eq:step3-1}
\end{equation}
Subsequently, ${\rm Alice_{2}}$ measures her qubit $S_{2}$ in $\{v_{0},v_{1}\}$
basis, where basis elements are already described in Step 3. The reduced
state at the end of the three receivers can be deduced from Eq.
(\ref{eq:step3-1}).
\item [{Case~2.1:}] \textbf{The receivers decide that ${\rm Bob_{2}}$
will reconstruct the state}
\end{description}
\textbf{Step~4-1} When ${\rm Bob_{2}}$ is supposed to reconstruct
the unknown quantum state $|\xi\rangle$, the reduced quantum state
can be decomposed as

\begin{equation}
\begin{array}{lcl}
|\phi_{0}^{\prime}\rangle & = & \frac{1}{\sqrt{2}}e^{-i\phi}\left\{ |00\rangle_{R_{1}R_{3}}(be^{i\phi}|0\rangle-a|1\rangle)_{R_{2}}+|11\rangle_{R_{1}R_{3}}(be^{i\phi}|1\rangle-a|0\rangle)_{R_{2}}\right\} ,\\
|\phi_{1}^{\prime}\rangle & = & \frac{1}{\sqrt{2}}\left\{ |00\rangle_{R_{1}R_{3}}(be^{i\phi}|0\rangle+a|1\rangle)_{R_{2}}+|11\rangle_{R_{1}R_{3}}(be^{i\phi}|1\rangle+a|0\rangle)_{R_{2}}\right\} ,
\end{array}\label{eq:step4-1}
\end{equation}
where the global phase $e^{-i\phi}$ can be ignored,\textcolor{magenta}{{}
}which is consistent with the quantum mechanics.

From Eq. (\ref{eq:step4-1}), it is clear that ${\rm Bob_{2}}$ can
reconstruct the quantum state $|\xi\rangle$ by applying suitable
Pauli operations with the help of measurement results of either ${\rm Bob_{1}}$
or ${\rm Bob_{3}}$ as both of them have the same measurement outcomes
(cf. Column 3 of Table \ref{tab:Relation-between-the-2}). Consequently,
the receiver ${\rm R_{2}}$ is accredited the higher power agent in
the proposed hierarchical scheme.

\begin{table}[H]
\begin{centering}
\begin{tabular}{|>{\centering}p{3cm}|>{\centering}p{3cm}|>{\centering}p{3cm}|>{\centering}p{3cm}|}
\hline 
${\rm Alice_{1}}$'s measurement outcome in $\{u_{0},u_{1}\}$ basis  & ${\rm Alice_{2}}$'s measurement outcome in $\{v_{0},v_{1}\}$ basis  & ${\rm Bob_{i}}$'s measurement outcome (in
$\{|0\rangle,|1\rangle\}$ basis), where $i\in\left\{1,3\right\}$  & Pauli operations ${\rm Bob_{2}}$ have to apply\tabularnewline
\hline 
 & $|v_{0}\rangle$  & $|0\rangle$  & $iY$\tabularnewline
\cline{3-4} 
$|u_{1}\rangle$  &  & $|1\rangle$  & $Z$ \tabularnewline
\cline{2-4} 
 & $|v_{1}\rangle$  & $|0\rangle$  & $X$\tabularnewline
\cline{3-4} 
 &  & $|1\rangle$  & $I$\tabularnewline
\hline 
\end{tabular}
\par\end{centering}

\protect\caption{\label{tab:Relation-between-the-2}When ${\rm Alice_{1}}$'s measurement
outcome is $|u_{1}\rangle$, corresponding measurement results of all
the remaining parties and the unitary operations to be applied
by ${\rm Bob_{2}}$ are summarized here. The receiver ${\rm Bob_{2}}$ needs the collaboration
of either of the agents ${\rm Bob_{1}}$ or ${\rm Bob_{3}}$.}
\end{table}

\begin{description}
\item [{Case~2.2:}] \textbf{The receivers decide that ${\rm Bob_{3}}$
will reconstruct the state}
\end{description}
\textbf{Step~$4-1^{\prime}$ }When ${\rm Bob_{3}}$ has to reconstruct
the quantum state, the reduced quantum state can be written (corresponding
to measurement outcomes of ${\rm Alice_{2}}$) as follows:

If the measurement outcome of ${\rm Alice_{2}}$ is $|v_{0}\rangle$,
the reduced state becomes

\begin{equation}
\begin{array}{lcl}
|\phi_{0}^{\prime}\rangle & = & \frac{e^{-i\phi}}{2}\left[\left(|+\rangle|0\rangle\right)_{R_{1}R_{2}}\left(be^{i\phi}|0\rangle-a|1\rangle\right)_{R_{3}}+\left(|-\rangle|0\rangle\right)_{R_{1}R_{2}}\left(be^{i\phi}|0\rangle+a|1\rangle\right)_{R_{3}}\right.\\
 & + & \left.\left(|+\rangle|1\rangle\right)_{R_{1}R_{2}}\left(be^{i\phi}|1\rangle-a|0\rangle\right)_{R_{3}}-\left(|-\rangle|1\rangle\right)_{R_{1}R_{2}}\left(be^{i\phi}|1\rangle+a|0\rangle\right)_{R_{3}}\right],
\end{array}\label{eq:step41-1}
\end{equation}
and if the measurement outcome of the sender ${\rm Alice_{2}}$ is
$|v_{1}\rangle$, the state can be written as 

\begin{equation}
\begin{array}{lcl}
|\phi_{1}^{\prime}\rangle & = & \frac{1}{2}\left[\left(|+\rangle|0\rangle\right)_{R_{1}R_{2}}\left(be^{i\phi}|0\rangle+a|1\rangle\right)_{R_{3}}+\left(|-\rangle|0\rangle\right)_{R_{1}R_{2}}\left(be^{i\phi}|0\rangle-a|1\rangle\right)_{R_{3}}\right.\\
 & + & \left.\left(|+\rangle|1\rangle\right)_{R_{1}R_{2}}\left(be^{i\phi}|1\rangle+a|0\rangle\right)_{R_{3}}-\left(|-\rangle|1\rangle\right)_{R_{1}R_{2}}\left(be^{i\phi}|1\rangle-a|0\rangle\right)_{R_{3}}\right].
\end{array}\label{eq:step41-2}
\end{equation}

From Eqs. (\ref{eq:step41-1}) and (\ref{eq:step41-2}), it can be
inferred that ${\rm Bob_{3}}$ can reconstruct $|\xi\rangle$ by applying
an appropriate Pauli operation only if both the remaining receivers
help him by conveying their measurement results after measuring their
qubits in suitable basis. In Table \ref{tab:Relation-between-the-1-1},
it has been explicitly shown that ${\rm Bob_{1}}$ and ${\rm Bob_{2}}$
measure their qubits in $\{|+\rangle,|-\rangle\}$ and $\{|0\rangle,|1\rangle\}$
bases, respectively. In the table, Pauli operations corresponding
to each possible measurement outcome are also clearly mentioned. Hence,
${\rm Bob_{3}}$ is the lower power agent in the proposed HJRSP scheme.
Interestingly, a similar situation is obtained for ${\rm Bob_{1}}$,
consequently he is at the same level of hierarchy as ${\rm Bob_{3}}$.

\begin{table}[H]
\begin{centering}
\begin{tabular}{|>{\centering}p{3cm}|>{\centering}p{3cm}|>{\centering}p{3cm}|>{\centering}p{3cm}|}
\hline 
${\rm Alice_{1}}$'s measurement outcome in $\{u_{0},u_{1}\}$ basis  & ${\rm Alice_{2}}$'s measurement outcome in $\{v_{0},v_{1}\}$ basis  & ${\rm Bob_{1}}$'s and ${\rm Bob_{2}}$'s joint measurement outcome
(in $\{|+\rangle,|-\rangle\}$,$\{|0\rangle,|1\rangle\}$ bases, respectively) & Pauli operation to be applied by ${\rm Bob_{3}}$\tabularnewline
\hline 
 &  & $|+\rangle|0\rangle$  & $iY$\tabularnewline
\cline{3-4} 
 & $|v_{0}\rangle$  & $|-\rangle|0\rangle$  & $X$\tabularnewline
\cline{3-4} 
 &  & $|+\rangle|1\rangle$  & $Z$\tabularnewline
\cline{3-4} 
$|u_{1}\rangle$  &  & $|-\rangle|1\rangle$  & $I$\tabularnewline
\cline{2-4} 
 &  & $|+\rangle|0\rangle$  & $X$\tabularnewline
\cline{3-4} 
 & $|v_{1}\rangle$  & $|-\rangle|0\rangle$  & $iY$\tabularnewline
\cline{3-4} 
 &  & $|+\rangle|1\rangle$  & $I$\tabularnewline
\cline{3-4} 
 &  & $|-\rangle|1\rangle$  & $Z$\tabularnewline
\hline 
\end{tabular}
\par\end{centering}

\protect\caption{\label{tab:Relation-between-the-1-1}The measurement outcomes of both
the senders and receivers are summarized with the Pauli operations that
${\rm Bob_{3}}$ needs to apply. The receiver ${\rm Bob_{3}}$ needs
the joint collaboration of all other agents.}
\end{table}

\subsection{Probabilistic HJRSP\label{sub:Probabilistic-HJRSP}}

In the last subsection, we have proposed a deterministic HJRSP scheme
using maximally entangled 5-qubit cluster state. In this section,
we aim to propose a scheme for probabilistic HJRSP using non-maximally
entangled 5-qubit cluster state of a specific form. To do so, we assume
that ${\rm Alice_{1}}$ prepares and shares a non-maximally entangled
5-qubit cluster state of the form

\begin{equation}
\begin{array}{lcl}
|C^{\prime}\rangle & = & \frac{1}{\sqrt{2}}\left[\alpha(|00000\text{\ensuremath{\rangle+|00111\rangle)+\beta(|11010\rangle+|11101\rangle)}}\right]_{S_{1}S_{2}R_{1}R_{2}R_{3}}\\
 & = & \frac{1}{\sqrt{2}}\left[|u_{0}\rangle_{S_{1}}\{\alpha a(|0000\text{\ensuremath{\rangle+|0111\rangle)+\beta b(|1010\rangle+|1101\rangle)\}}}\right.\\
 & + & \left.|u_{1}\rangle_{S_{1}}\{\alpha b(|0000\text{\ensuremath{\rangle+|0111\rangle)-\beta a(|1010\rangle+|1101\rangle)}}\}\right]_{S_{2}R_{1}R_{2}R_{3}},
\end{array}\label{eq:prob-step-1}
\end{equation}
where $|\alpha|^{2}+|\beta|^{2}=1$ with $|\alpha|\neq\frac{1}{\sqrt{2}}$
and $|u_{0}\rangle$ and $|u_{1}\rangle$ are already defined in the
context of the previous protocol. From Eq. (\ref{eq:prob-step-1})
it can be observed that if ${\rm Alice_{1}}$ measures her qubit $S_{1}$
in $\left\{ |u_{0}\rangle,|u_{1}\rangle\right\} $ basis, then the
reduced quantum state corresponding to measurement outcome $|u_{0}\rangle$
and $|u_{1}\rangle$ are given as 

\begin{equation}
\begin{array}{lcl}
|\psi_{0}\rangle & = & \frac{1}{\sqrt{2\left(a^{2}|\alpha|^{2}+b^{2}|\beta|^{2}\right)}}\left[\alpha a(|0000\text{\ensuremath{\rangle+|0111\rangle)+\beta b(|1010\rangle+|1101\rangle)}}\right]_{S_{2}R_{1}R_{2}R_{3}},\\
|\psi_{1}\rangle & = & \frac{1}{\sqrt{2\left(b^{2}|\alpha|^{2}+a^{2}|\beta|^{2}\right)}}\left[\alpha b(|0000\text{\ensuremath{\rangle+|0111\rangle)-\beta a(|1010\rangle+|1101\rangle)}}\right]_{S_{2}R_{1}R_{2}R_{3}},
\end{array}\label{eq:prob-alice1}
\end{equation}
respectively.
\begin{description}
\item [{Case~1:}] \textbf{${\rm Alice_{1}}$'s measurement outcome is
$|u_{0}\rangle$ }
\end{description}
Then ${\rm Alice_{2}}$ applies a phase operator $P\left(2\phi\right)=\left[\begin{array}{lc}
1 & 0\\
0 & e^{2i\phi}
\end{array}\right]$ on $\text{|}\psi_{0}\rangle$, as in the deterministic HJRSP scheme
defined above, and the state can be written as

\begin{equation}
\begin{array}{lcc}
\text{|}\psi_{0}^{\prime}\rangle & = & \left(P\left(2\phi\right)\otimes I^{\otimes3}\right)\text{|}\psi_{0}\rangle=\frac{1}{\sqrt{2\left(a^{2}|\alpha|^{2}+b^{2}|\beta|^{2}\right)}}[\alpha a(|0000\rangle+|0111\rangle)+\beta be^{2i\phi}(|1010\rangle+|1101\rangle)]_{S_{2}R_{1}R_{2}R_{3}}.\end{array}\label{eq:prob-step2}
\end{equation}

As the transformed state can be written as 
\begin{equation}
\begin{array}{lcl}
|\psi_{0}^{\text{\ensuremath{\prime}}}\rangle & = & \frac{1}{2\sqrt{\left(a^{2}|\alpha|^{2}+b^{2}|\beta|^{2}\right)}}\left[|v_{0}\rangle_{S_{2}}\{\alpha a(|000\rangle+|111\rangle)+\beta be^{i\phi}(|010\rangle+|101\rangle)\}_{R_{1}R_{2}R_{3}}\right.\\
 & + & e^{i\phi}\left.|v_{1}\rangle_{S_{2}}\{\alpha a(|000\rangle+|111\rangle)-\beta be^{i\phi}(|010\rangle+|101\rangle)\}_{R_{1}R_{2}R_{3}}\right]\\
 & = & \frac{1}{\sqrt{2}}\left[|v_{0}\rangle_{S_{2}}|\Phi_{0}\rangle_{R_{1}R_{2}R_{3}}+|v_{1}\rangle_{S_{2}}|\Phi_{1}\rangle_{R_{1}R_{2}R_{3}}\right],
\end{array}\label{eq:prob-step2-1}
\end{equation}
where $|\Phi_{0}\rangle=\frac{1}{\sqrt{2\left(a^{2}|\alpha|^{2}+b^{2}|\beta|^{2}\right)}}\{\alpha a(|000\rangle+|111\rangle)+\beta be^{i\phi}(|010\rangle+|101\rangle)\}_{R_{1}R_{2}R_{3}}$
and $|\Phi_{1}\rangle=\frac{1}{\sqrt{2\left(a^{2}|\alpha|^{2}+b^{2}|\beta|^{2}\right)}}\{\alpha a(|000\rangle+|111\rangle)-\beta be^{i\phi}(|010\rangle+|101\rangle)\}_{R_{1}R_{2}R_{3}}$.
Further, ${\rm Alice_{2}}$ measures her qubit $S_{2}$ in $\{v_{0},v_{1}\}$
basis, where $|v_{0}\rangle$ and $|v_{1}\rangle$ are as defined
in the previous subsection.
\begin{description}
\item [{Case~1.1:}] \textbf{The receivers decide that ${\rm Bob_{2}}$
will recover the state}
\end{description}
If the agents decide that ${\rm Bob_{2}}$ will reconstruct the unknown
quantum state $|\xi\rangle$, and other two receivers would measure
their qubits in computational basis, then the reduced quantum state
can be decomposed as

\begin{equation}
\begin{array}{lcl}
|\Phi_{0}\rangle & = & \frac{1}{\sqrt{2\left(a^{2}|\alpha|^{2}+b^{2}|\beta|^{2}\right)}}\{|00\rangle_{R_{1}R_{3}}(\alpha a|0\rangle+\beta be^{i\phi}|1\rangle)_{R_{2}}+|11\rangle_{R_{1}R_{3}}(\alpha a|1\rangle+\beta be^{i\phi}|0\rangle)_{R_{2}}\},\\
|\Phi_{1}\rangle & = & \frac{1}{\sqrt{2\left(a^{2}|\alpha|^{2}+b^{2}|\beta|^{2}\right)}}\{|00\rangle_{R_{1}R_{3}}(\alpha a|0\rangle-\beta be^{i\phi}|1\rangle)_{R_{2}}+|11\rangle_{R_{1}R_{3}}(\alpha a|1\rangle-\beta be^{i\phi}|0\rangle)_{R_{2}}\}.
\end{array}\label{eq:prob-step3}
\end{equation}

From Eq. (\ref{eq:prob-step3}), it is clear that ${\rm Bob_{2}}$
can not directly reconstruct the quantum state just by applying the
Pauli operators even if all the receivers cooperate. Therefore, he
has to change the strategy as follows: ${\rm Bob_{2}}$ prepares an
ancilla qubit in $|0\rangle_{{\rm aux}}$ and applies the following
2-qubit unitary operations $U_{0}/U_{1}$ on his qubits (i.e., on
the combined system of his existing qubit and ancilla), where 
\begin{equation}
U_{0}=\left(\begin{array}{cccc}
\frac{\beta}{\alpha} & \sqrt{1-\frac{\beta^{2}}{\alpha^{2}}} & 0 & 0\\
0 & 0 & 0 & -1\\
0 & 0 & 1 & 0\\
\sqrt{1-\frac{\beta^{2}}{\alpha^{2}}} & -\frac{\beta}{\alpha} & 0 & 0
\end{array}\right),\label{eq:U-probabilistic-HJRSP}
\end{equation}
and
\begin{equation}
U_{1}=U_{0}\left(X\otimes I\right)=\left(\begin{array}{cccc}
0 & 0 & \frac{\beta}{\alpha} & \sqrt{1-\frac{\beta^{2}}{\alpha^{2}}}\\
0 & -1 & 0 & 0\\
1 & 0 & 0 & 0\\
0 & 0 & \sqrt{1-\frac{\beta^{2}}{\alpha^{2}}} & -\frac{\beta}{\alpha}
\end{array}\right).\label{eq:U1-probabilistic-HJRSP}
\end{equation}
As $\alpha$ and $\beta$ are known, the construction of $U_{0}/U_{1}$
is possible. In Eq. (\ref{eq:prob-step3}), let us consider a case
that the measurement outcomes of ${\rm Alice_{2}}$ and ${\rm Bob_{1}}$
(or ${\rm Bob_{3}}$ as the measurement outcomes of both lower power
agents are the same) are $|v_{0}\rangle$ and $|0\rangle(|1\rangle)$,
respectively. Then ${\rm Bob_{2}}$ has to apply $U_{0}(U_{1})$ on
his product state. For instance, 
\[
|\chi\rangle_{1^{\prime}}=|\chi\rangle_{1}|0\rangle_{{\rm aux}}=\frac{1}{\sqrt{\left(a^{2}|\alpha|^{2}+b^{2}|\beta|^{2}\right)}}\left(\alpha a|0\rangle+\beta be^{i\phi}|1\rangle\right)|0\rangle_{{\rm aux}},
\]
where $|\chi\rangle_{1}$ is ${\rm Bob_{2}}$'s reduced quantum state
corresponding to ${\rm Bob_{1}}$'s measurement outcome $|0\rangle$.
Subsequent operation of $U_{0}$ on this composite state gives
\[
\begin{array}{lcl}
U_{0}|\chi\rangle_{1^{\prime}} & = & \frac{1}{\sqrt{\left(a^{2}|\alpha|^{2}+b^{2}|\beta|^{2}\right)}}\left\{ \beta(a|0\rangle+be^{i\phi}|1\rangle)|0\rangle+a\sqrt{\alpha^{2}-\beta^{2}}|1\rangle|1\rangle\right\} .\end{array}
\]
Finally, ${\rm Bob_{2}}$  measures the last qubit (ancilla) in
the computational basis $\{|0\rangle,|1\rangle\}$. If his measurement
yields $|0\rangle$ then he obtains unknown state with unit fidelity,
but if his measurement on ancilla yields $|1\rangle$ then he fails to reconstruct the state. In other words, he can recover the unknown state only when he obtains a specific measurement outcome. Thus, in analogy to the probabilistic teleportation
scheme we may refer to this scheme as probabilistic hierarchical joint
remote state preparation scheme. Similarly, we can check the other
three possibilities. All these results are summarized in the Table
\ref{tab:Relation-between-the-3}.

\begin{table}[H]
\begin{centering}
\begin{tabular}{|>{\centering}p{1.5cm}|>{\centering}p{1.5cm}|>{\centering}p{2.5cm}|>{\centering}p{2cm}|>{\centering}p{2cm}|}
\hline 
${\rm Alice_{1}}$'s measurement outcome in $\{u_{0},u_{1}\}$ basis  & ${\rm Alice_{2}}$'s measurement outcome in $\{v_{0},v_{1}\}$ basis  & ${\rm Bob_{i}}$'s measurement outcome (in
$\{|0\rangle,|1\rangle\}$ basis), where $i\in\left\{1,3\right\}$ & Two qubit unitary operation to be applied by ${\rm Bob_{2}}$ ($U_{0}/U_{1}$)  & Pauli operation ${\rm Bob_{2}}$ have to apply to obtain $|\xi\rangle$\tabularnewline
\hline 
 & $|v_{0}\rangle$  & $|0\rangle$  & $U_{0}$  & $I$ \tabularnewline
\cline{3-5} 
$|u_{0}\rangle$  &  & $|1\rangle$  & $U_{1}$  & $I$ \tabularnewline
\cline{2-5} 
 & $|v_{1}\rangle$  & $|0\rangle$  & $U_{0}$  & $Z$ \tabularnewline
\cline{3-5} 
 &  & $|1\rangle$  & $U_{1}$  & $Z$ \tabularnewline
\hline 
\end{tabular}
\par\end{centering}

\protect\caption{\label{tab:Relation-between-the-3}The probabilistic HJRSP scheme
is summarized here for all the successful cases. The measurement outcomes
of all the senders and receivers are summed up with the unitary and
Pauli operations ${\rm Bob_{2}}$ needs to apply. It can be deduced
from the third column that the receiver ${\rm Bob_{2}}$ is a higher
power agent as he needs the collaboration of either of the agents.}
\end{table}

\begin{description}
\item [{Case~1.2:}] \textbf{The receivers decide that ${\rm Bob_{3}}$
will recover the state}
\end{description}
If the agents unanimously agreed that ${\rm Bob_{3}}$ will reconstruct
the unknown quantum state, the reduced quantum state of three receivers
can be decomposed as follows. 

When the measurement outcome of ${\rm Alice_{2}}$ is $|v_{0}\rangle$ 

\begin{equation}
\begin{array}{lcl}
|\Phi_{0}\rangle & = & \frac{1}{2\sqrt{\left(a^{2}|\alpha|^{2}+b^{2}|\beta|^{2}\right)}}\left[\left(|+\rangle|0\rangle\right)_{R_{1}R_{2}}\left(\alpha a|0\rangle+\beta be^{i\phi}|1\rangle\right)_{R_{3}}+\left(|-\rangle|0\rangle\right)_{R_{1}R_{2}}\left(\alpha a|0\rangle-\beta be^{i\phi}|1\rangle\right)_{R_{3}}\right.\\
 & + & \left.\left(|+\rangle|1\rangle\right)_{R_{1}R_{2}}\left(\alpha a|1\rangle+\beta be^{i\phi}|0\rangle\right)_{R_{3}}-\left(|-\rangle|1\rangle\right)_{R_{1}R_{2}}\left(\alpha a|1\rangle-\beta be^{i\phi}|0\rangle\right)_{R_{3}}\right],
\end{array}\label{eq:prob-step5}
\end{equation}

When the measurement outcome of ${\rm Alice_{2}}$ is $|v_{1}\rangle$ 

\begin{equation}
\begin{array}{lcl}
|\Phi_{1}\rangle & = & \frac{1}{2\sqrt{\left(a^{2}|\alpha|^{2}+b^{2}|\beta|^{2}\right)}}\left[\left(|+\rangle|0\rangle\right)_{R_{1}R_{2}}\left(\alpha a|0\rangle-\beta be^{i\phi}|1\rangle\right)_{R_{3}}+\left(|-\rangle|0\rangle\right)_{R_{1}R_{2}}\left(\alpha a|0\rangle+\beta be^{i\phi}|1\rangle\right)_{R_{3}}\right.\\
 & + & \left.\left(|+\rangle|1\rangle\right)_{R_{1}R_{2}}\left(\alpha a|1\rangle-\beta be^{i\phi}|0\rangle\right)_{R_{3}}-\left(|-\rangle|1\rangle\right)_{R_{1}R_{2}}\left(\alpha a|1\rangle+\beta be^{i\phi}|0\rangle\right)_{R_{3}}\right].
\end{array}\label{eq:prob-step5-1}
\end{equation}

From Eqs. (\ref{eq:prob-step5}) and (\ref{eq:prob-step5-1}), it
is clear that ${\rm Bob_{3}}$ can not directly reconstruct the quantum
state as in the previous case. Hence, similar to the strategy followed
by ${\rm Bob_{2}}$ above, ${\rm Bob_{3}}$ also applies a unitary
operation $U_{0}/U_{1}$ on his composite system (i.e., on his existing
qubit and ancilla). Finally, he can reconstruct the quantum state
by further application of an appropriate Pauli operation as shown
in Table \ref{tab:Relation-between-the-4}.

\begin{table}[H]
\begin{centering}
\begin{tabular}{|>{\centering}p{1.5cm}|>{\centering}p{1.5cm}|>{\centering}p{2.5cm}|>{\centering}p{2.5cm}|>{\centering}p{2cm}|>{\centering}p{1.7cm}|}
\hline 
${\rm Alice_{1}}$'s measurement outcome in $\{u_{0},u_{1}\}$ basis  & ${\rm Alice_{2}}$'s measurement outcome in $\{v_{0},v_{1}\}$ basis  & ${\rm Bob_{1}}$'s measurement outcome (in $\{|+\rangle,|-\rangle\}$
basis) & ${\rm Bob_{2}}$'s measurement outcomes (in $\{|0\rangle,|1\rangle\}$
basis) & Two qubit unitary operation to be applied by ${\rm Bob_{3}}$ ($U_{0}/U_{1}$)  & Pauli operation to be applied by ${\rm Bob_{3}}$ to reconstruct $|\xi\rangle$\tabularnewline
\hline 
 &  & $|+\rangle$  & $|0\rangle$  & $U_{0}$  & $I$ \tabularnewline
\cline{3-6} 
 & $|v_{0}\rangle$  & $|-\rangle$  & $|0\rangle$  & $U_{0}$  & $Z$ \tabularnewline
\cline{3-6} 
 &  & $|+\rangle$  & $|1\rangle$  & $U_{1}$  & $I$ \tabularnewline
\cline{3-6} 
$|u_{0}\rangle$  &  & $|-\rangle$  & $|1\rangle$  & $U_{1}$  & $Z$ \tabularnewline
\cline{2-6} 
 &  & $|+\rangle$  & $|0\rangle$  & $U_{0}$  & $Z$ \tabularnewline
\cline{3-6} 
 & $|v_{1}\rangle$  & $|-\rangle$  & $|0\rangle$  & $U_{0}$  & $I$ \tabularnewline
\cline{3-6} 
 &  & $|+\rangle$  & $|1\rangle$  & $U_{1}$  & $Z$ \tabularnewline
\cline{3-6} 
 &  & $|-\rangle$  & $|1\rangle$  & $U_{1}$  & $I$ \tabularnewline
\hline 
\end{tabular}
\par\end{centering}

\protect\caption{\label{tab:Relation-between-the-4}All possible successful cases in
the probabilistic HJRSP scheme are summarized with the measurement
outcomes of all the senders and receivers with corresponding unitary
and Pauli operations to be implemented by ${\rm Bob_{3}}$ to recover
the quantum state. The receiver ${\rm Bob_{3}}$ needs the joint collaboration
of all the other agents.}
\end{table}

\begin{description}
\item [{Case~2:}] \textbf{${\rm Alice_{1}}$'s measurement outcome is
$|u_{1}\rangle$ }
\end{description}
Then ${\rm Alice_{2}}$ need not apply a phase operator $P\left(2\phi\right)$
on $\text{|}\psi_{1}\rangle$ in Eq. (\ref{eq:prob-alice1}). Subsequently,
${\rm Alice_{2}}$ measures her qubit $S_{2}$ in $\{v_{0},v_{1}\}$
basis, where $|v_{0}\rangle$ and $|v_{1}\rangle$ have the same meaning
as in the last subsection. As the state can be written as

\begin{equation}
\begin{array}{lcl}
|\psi_{1}^{\text{\ensuremath{\prime\prime}}}\rangle & = & \frac{1}{2\sqrt{\left(b^{2}|\alpha|^{2}+a^{2}|\beta|^{2}\right)}}\left[|v_{0}\rangle_{S_{2}}\{\alpha b(|000\rangle+|111\rangle)-\beta ae^{-i\phi}(|010\rangle+|101\rangle)\}_{R_{1}R_{2}R_{3}}\right.\\
 & + & \left.|v_{1}\rangle_{S_{2}}\{e^{i\phi}\alpha b(|000\rangle+|111\rangle)+\beta a(|010\rangle+|101\rangle)\}_{R_{1}R_{2}R_{3}}\right],\\
 & = & \frac{1}{\sqrt{2}}\left[|v_{0}\rangle_{S_{2}}|\Phi_{0}^{\prime}\rangle_{R_{1}R_{2}R_{3}}+|v_{1}\rangle_{S_{2}}|\Phi_{1}^{\prime}\rangle_{R_{1}R_{2}R_{3}}\right],
\end{array}\label{eq:prob-case-2-1}
\end{equation}
from which the quantum state after measurement can be deduced to be
$|\Phi_{0}^{\prime}\rangle$ ($|\Phi_{1}^{\prime}\rangle$) for ${\rm Alice_{2}}$'s
measurement outcomes $|v_{0}\rangle$ ($|v_{1}\rangle$).
\begin{description}
\item [{Case~2.1:}] \textbf{The receivers decide that ${\rm Bob_{2}}$
will recover the state}
\end{description}
If all the agents decree ${\rm Bob_{2}}$ to reconstruct the unknown
quantum state, the reduced quantum state of the receivers can be decomposed
as

\begin{equation}
\begin{array}{lcl}
|\Phi_{0}^{\prime}\rangle & = & \frac{1}{\sqrt{2\left(b^{2}|\alpha|^{2}+a^{2}|\beta|^{2}\right)}}\left\{ |00\rangle_{R_{1}R_{3}}(\alpha b|0\rangle-\beta ae^{-i\phi}|1\rangle)_{R_{2}}+|11\rangle_{R_{1}R_{3}}(\alpha b|1\rangle-\beta ae^{-i\phi}|0\rangle)_{R_{2}}\right\} ,\\
|\Phi_{1}^{\prime}\rangle & = & \frac{1}{\sqrt{2\left(b^{2}|\alpha|^{2}+a^{2}|\beta|^{2}\right)}}\left\{ |00\rangle_{R_{1}R_{3}}(\alpha be^{i\phi}|0\rangle+\beta a|1\rangle)_{R_{2}}+|11\rangle_{R_{1}R_{3}}(\alpha be^{i\phi}|1\rangle+\beta a|0\rangle)_{R_{2}}\right\} .
\end{array}\label{eq:prob-case2}
\end{equation}
From Eq. (\ref{eq:prob-case2}), it is concluded that ${\rm Bob_{2}}$
can not directly reconstruct the quantum state unless he adopts the
same strategy as was discussed for the case of ${\rm Alice_{1}}$'s
measurement outcome $|u_{0}\rangle$. In analogy of Case 1, when ${\rm Alice_{1}}$
obtained $|u_{0}\rangle$, specific cases may be studied. Here, we
have restricted ourselves only to the successful cases of probabilistic
HJRSP which summarized in Table \ref{tab:Relation-between-the-5}.\textcolor{blue}{{} }

\begin{table}[H]
\begin{centering}
\begin{tabular}{|>{\centering}p{1.5cm}|>{\centering}p{1.5cm}|>{\centering}p{2.5cm}|>{\centering}p{2cm}|>{\centering}p{2cm}|}
\hline 
${\rm Alice_{1}}$'s measurement outcome in $\{u_{0},u_{1}\}$ basis  & ${\rm Alice_{2}}$'s measurement outcome in $\{v_{0},v_{1}\}$ basis  & ${\rm Bob_{i}}$'s measurement outcome (in
$\{|0\rangle,|1\rangle\}$ basis), where $i\in\left\{1,3\right\}$  & Two qubit unitary operation to be applied by ${\rm Bob_{2}}$ ($U_{0}/U_{1}$)  & Pauli operation ${\rm Bob_{2}}$ have to apply to reconstruct $|\xi\rangle$\tabularnewline
\hline 
 & $|v_{0}\rangle$  & $|0\rangle$  & $U_{0}$  & $iY$ \tabularnewline
\cline{3-5} 
$|u_{1}\rangle$  &  & $|1\rangle$  & $U_{1}$  & $iY$ \tabularnewline
\cline{2-5} 
 & $|v_{1}\rangle$  & $|0\rangle$  & $U_{0}$  & $X$ \tabularnewline
\cline{3-5} 
 &  & $|1\rangle$  & $U_{1}$  & $X$ \tabularnewline
\hline 
\end{tabular}
\par\end{centering}

\protect\caption{\label{tab:Relation-between-the-5}All successful cases of probabilistic
HJRSP are summarized with the measurement outcomes of all the senders
and receivers with corresponding unitary and Pauli operations which
${\rm Bob_{2}}$ applies to recover the quantum state. The receiver
${\rm Bob_{2}}$ remains a higher power agent as in deterministic
HJRSP scheme.}
\end{table}

\textbf{Case~2.2: The receivers decide that ${\rm Bob_{3}}$ will
recover the state}

When ${\rm Bob_{3}}$ is assigned the task to reconstruct the unknown
quantum state, the reduced quantum state can be decomposed as follows. 

When the measurement outcome of ${\rm Alice_{2}}$ is $|v_{0}\rangle$ 

\begin{equation}
\begin{array}{lcl}
|\Phi_{0}^{\prime}\rangle & = & \frac{1}{2\sqrt{\left(b^{2}|\alpha|^{2}+a^{2}|\beta|^{2}\right)}}\left[\left(|+\rangle|0\rangle\right)_{R_{1}R_{2}}\left(\alpha b|0\rangle-\beta ae^{-i\phi}|1\rangle\right)_{R_{3}}+\left(|-\rangle|0\rangle\right)_{R_{1}R_{2}}\left(\alpha b|0\rangle+\beta ae^{-i\phi}|1\rangle\right)_{R_{3}}\right.\\
 & + & \left.\left(|+\rangle|1\rangle\right)_{R_{1}R_{2}}\left(\alpha b|1\rangle-\beta ae^{-i\phi}|0\rangle\right)_{R_{3}}-\left(|-\rangle|1\rangle\right)_{R_{1}R_{2}}\left(\alpha b|1\rangle+\beta ae^{-i\phi}|0\rangle\right)_{R_{3}}\right],
\end{array}\label{eq:prob-case2-2}
\end{equation}

When ${\rm Alice_{2}}$ obtains $|v_{1}\rangle$ 

\begin{equation}
\begin{array}{lcl}
|\Phi_{1}^{\prime}\rangle & = & \frac{1}{2\sqrt{\left(b^{2}|\alpha|^{2}+a^{2}|\beta|^{2}\right)}}\left[\left(|+\rangle|0\rangle\right)_{R_{1}R_{2}}\left(\alpha be^{i\phi}|0\rangle+\beta a|1\rangle\right)_{R_{3}}+\left(|-\rangle|0\rangle\right)_{R_{1}R_{2}}\left(\alpha be^{i\phi}|0\rangle-\beta a|1\rangle\right)_{R_{3}}\right.\\
 & + & \left.\left(|+\rangle|1\rangle\right)_{R_{1}R_{2}}\left(\alpha be^{i\phi}|1\rangle+\beta a|0\rangle\right)_{R_{3}}-\left(|-\rangle|1\rangle\right)_{R_{1}R_{2}}\left(\alpha be^{i\phi}|1\rangle-\beta a|0\rangle\right)_{R_{3}}\right].
\end{array}\label{eq:prob-case2-3}
\end{equation}
Similar to the earlier cases of probabilistic HJRSP, ${\rm Bob_{3}}$
will have to apply one of the unitary operations $U_{0}$ or $U_{1}$
before reconstructing the state by operating a suitable Pauli gate,
which depends on the measurement outputs of ${\rm Bob_{1}}$ and ${\rm Bob_{2}}$.
All the successful cases of probabilistic HJRSP are entabulated in
Table \ref{tab:Relation-between-the-6}.

\begin{table}[H]
\begin{centering}
\begin{tabular}{|>{\centering}p{1.5cm}|>{\centering}p{1.5cm}|>{\centering}p{2.5cm}|>{\centering}p{2.5cm}|>{\centering}p{2cm}|>{\centering}p{1.7cm}|}
\hline 
${\rm Alice_{1}}$'s measurement outcome in $\{u_{0},u_{1}\}$ basis  & ${\rm Alice_{2}}$'s measurement outcome in $\{v_{0},v_{1}\}$ basis  & ${\rm Bob_{1}}$'s measurement outcome (in $\{|+\rangle,|-\rangle\}$
basis) & ${\rm Bob_{2}}$'s measurement outcome (in $\{|0\rangle,|1\rangle\}$
basis) & Two qubit unitary operation to be applied by ${\rm Bob_{3}}$ ($U_{0}/U_{1}$)  & Pauli operation to be applied by ${\rm Bob_{3}}$ to reconstruct $|\xi\rangle$\tabularnewline
\hline 
 & $|v_{0}\rangle$  & $|+\rangle$  & $|0\rangle$  & $U_{0}$  & $iY$ \tabularnewline
\cline{3-6} 
 &  & $|-\rangle$  & $|0\rangle$  & $U_{0}$  & $X$ \tabularnewline
\cline{3-6} 
 &  & $|+\rangle$  & $|1\rangle$  & $U_{1}$  & $iY$ \tabularnewline
\cline{3-6} 
$|u_{1}\rangle$  &  & $|-\rangle$  & $|1\rangle$  & $U_{1}$  & $X$ \tabularnewline
\cline{2-6} 
 & $|v_{1}\rangle$  & $|+\rangle$  & $|0\rangle$  & $U_{0}$  & $X$ \tabularnewline
\cline{3-6} 
 &  & $|-\rangle$  & $|0\rangle$  & $U_{0}$  & $iY$ \tabularnewline
\cline{3-6} 
 &  & $|+\rangle$  & $|1\rangle$  & $U_{1}$  & $X$ \tabularnewline
\cline{3-6} 
 &  & $|-\rangle$  & $|1\rangle$  & $U_{1}$  & $iY$ \tabularnewline
\hline 
\end{tabular}
\par\end{centering}

\protect\caption{\label{tab:Relation-between-the-6}All the successful cases of Probabilistic
HJRSP scheme when ${\rm Alice_{1}}$'s measurement result is $|u_{1}\rangle$
are listed here. The unitary operation and suitable Pauli gates ${\rm Bob_{3}}$
will require are mentioned against all possible measurement outcomes
of the senders and receivers. ${\rm Bob_{3}}$ can be seen the lowest
power agent here as well. }
\end{table}

\section{Practical Applications\label{sec:Practical-Applications}}

Hierarchical communications play important roles in our day to day
life. Several examples of practical situations where hierarchical
communication is essential have been discussed in our earlier works
\cite{Shukla,HDQSS}. Specifically, its relevance is discussed in
context of HQSS \cite{Shukla}, HQIS \cite{hierarchical}, and HDQSS
\cite{HDQSS}. In all those cases, the information or quantum state
to be shared was in possession of a single person (whom we referred
to as Alice or Sender). In contrast, here the initial state is jointly
possessed by two senders. In many practical scenarios, it allows an
avenue for joint decision, and thus, reduces risk associated with
policy decisions taken by a single person. Let us elaborate this particular
feature through a specific example. 

Consider that there exists a code to unlock a
nuclear weapon. Because of the fatal effect it may cause, this particular
code cannot be given to a single authorized person. Thus, the information (code)
has to be distributed among at least two authorized persons, so that
none of the authorized persons can misuse the code. Now, consider
that ${\rm Alice_{1}}$ and ${\rm Alice_{2}}$ (two authorized persons)
are the Prime Minister and President of a country, and ${\rm Bob}_{2}$
is the defense minister, ${\rm Bob}_{1}$ and ${\rm Bob}_{3}$ are
the defense secretary and the chief of the armed forces of that country,
respectively. If and only if the Prime Minister and President together
wish to permit the use of the nuclear weapon at a suitable time, then
they jointly distribute the information (the code which is required
to unlock the nuclear weapon) among the defense minister, the defense
secretary and the chief of the armed forces in such a way that the
minister can unlock the weapon if either the defense secretary or
the chief of the armed forces agrees and cooperates with him. However,
if the chief of the armed forces or the defense secretary wants to
unlock the weapon they would require the cooperation of each other
and that of the defense minister. Thus, the defense minister is more
powerful than the chief of the armed forces and the defense secretary,
but even he is not powerful enough to unlock the weapon alone, and
the senders (i.e., the Prime Minister and the President) are not powerful
enough to issue an individual order that allows the receivers to unlock
the weapon. This type of joint responsibility, is essential and routinely
exercised in a democracy. However, earlier proposed hierarchical schemes
of quantum communication did not contain this particular feature.
We can discuss a large number of similar practical situations where
HJRSP or a variant of it is essential, but it's not our purpose to
provide a long list of practical situation. Rather, we are interested
in analyzing the robustness and efficiency of the deterministic HJRSP
scheme proposed here.

\section{Effect of a set of noise models on the HJRSP scheme\label{sec:Effect-of-set}}

In the recent past, several schemes for classical and quantum communication
tasks have been theoretically proposed in the ideal situations, i.e.,
without considering the effects of noise present in the communication
channel. However, it's well understood that in any practical situation,
noise would play a crucial role, and the success of a scheme would
depend on the noise present in the channel. This fact motivates us
to investigate the effect of different quantum noise models on the
HJRSP schemes proposed here. Specifically, in this section we will
investigate the effect of AD noise, PD noise, collective dephasing
(CD) noise, collective rotation (CR) noise and Pauli noise. 

In this section, we aim to analyze the feasibility of the implementation
of the proposed deterministic HJRSP scheme in a noisy environment.
To quantitatively investigate the effect of noise on a scheme of quantum
communication, fidelity, which is a distance based measure, is usually
used. Specifically, the fidelity of the state obtained after considering
the effect of noise with the reconstructed state in the ideal case
can be obtained as 
\begin{equation}
F=\langle T|\rho_{k}|T\rangle.\label{eq:fdly}
\end{equation}
Here, $|T\rangle$ is the quantum state reconstructed after implementing
the HJRSP scheme in an ideal situation, while $\rho_{k}$ is the density
matrix of the quantum state obtained after considering the effect
of an interaction with the surrounding (i.e., when noise is present).
To be precise, the definition of fidelity used here (also used in
in Refs. \cite{CBRSP,RSP-with-noise,Kishore-decoy,crypt-switch,fdly})
is slightly different from the conventional one, $F^{\prime}(\sigma,\rho)=Tr\sqrt{\sigma^{\frac{1}{2}}\rho\sigma^{\frac{1}{2}}}$.
Further, in what follows, we will use the strategy adopted in Refs.
\cite{CBRSP,RSP-with-noise,Kishore-decoy,crypt-switch,fdly} to study
the effect of noise. It is a reasonable assumption that the qubits
not traveling through the channel are hardly affected by the noisy
environment. Hence, we have not considered the effect of noise on
the home qubit, which ${\rm Alice_{1}}$ has prepared and kept for
herself. Consider the initial quantum state $\rho=|C\rangle_{S_{1}S_{2}R_{1}R_{2}R_{3}\,S_{1}S_{2}R_{1}R_{2}R_{3}}\langle C|$,
where $|C\rangle_{S_{1}S_{2}R_{1}R_{2}R_{3}}$ is the cluster state
given in Eq. (\ref{eq:state}). The transformed density matrix under
the effect of AD or PD noisy channel can be expressed as 
\begin{equation}
\rho_{k}=\sum_{i,j,k,l}I_{2,S_{1}}\otimes E_{i,S_{2}}^{k}\otimes E_{j,R_{1}}^{k}\otimes E_{k,R_{2}}^{k}\otimes E_{l,R_{3}}^{k}\rho\left(I_{2,S_{1}}\otimes E_{i,S_{2}}^{k}\otimes E_{jR_{1}}^{k}\otimes E_{k,R_{2}}^{k}\otimes E_{l,R_{3}}^{k}\right)^{\dagger},\label{eq:After noise}
\end{equation}
where $I_{2}$ is a $2\times2$ identity matrix, and its application
on qubit $S_{1}$ corresponds to unaffected home qubit of ${\rm Alice_{1}}$.
For the remaining qubits $E_{J}^{k}$ are the Kraus operators for
AD or PD noise channels with $k\in\{AD,PD\}$ for AD and PD noise,
respectively. Here, $J\in\{0,1\}$ for AD and $J\in\{0,1,2\}$ for
PD noise models. The Kraus operators for AD and PD noise channels
will be described in detail in the following subsections. In the subscripts
of the Kraus operators the qubit on which it operates are also mentioned.
The same strategy may be used for the investigation of the effect
of Pauli channels, where the four Pauli gates (including identity
operator) are used to study the errors introduced due to noisy channel.
Further, the evolved quantum state under the collective noise models
can be described as 
\begin{equation}
\rho_{k}=\left(I_{2,S_{1}}\otimes U_{k}^{\otimes4}\right)\rho\left(I_{2,S_{1}}\otimes U_{k}^{\otimes4}\right)^{\dagger},\label{eq:under collection noise}
\end{equation}
where the subscript $k$ is $CD$ or $CR$ for CD or CR noise channels,
and $U_{k}$ is a $2\times2$ unitary matrix for either CD or CR noise
channels.

In the following subsections, we will analyze the deterministic HJRSP
scheme subjected to various noise models after briefly introducing
them. Further, we will discuss the dependence of the obtained fidelity
expressions on the noise parameters, which quantitatively illustrates
the effect of noise on the scheme. Here, we will refrain from considering
the effect of noise on the probabilistic HJRSP scheme, which will
be discussed elsewhere.

\subsection{Amplitude damping (AD) noise:}

The AD noise model is represented by the following Kraus operators
\cite{nielsen,RSP-with-noise,preskill} 
\begin{equation}
E_{0}^{AD}=\left[\begin{array}{cc}
1 & 0\\
0 & \sqrt{1-\eta_{A}}
\end{array}\right],\,\,\,\,\,\,\,\,\,\,\,\,\,\,\,E_{1}^{AD}=\left[\begin{array}{cc}
0 & \sqrt{\eta_{A}}\\
0 & 0
\end{array}\right],\label{eq:Krauss-amp-damping}
\end{equation}
where $\eta_{A}$ ($0\leq\eta_{A}\leq1$) is the decoherence rate
and is the probability of energy loss when a travel qubit passes through
an AD channel. Specifically, an AD channel simulates the interaction
of a quantum system with a vacuum bath. Using Eqs. (\ref{eq:fdly}),
(\ref{eq:After noise}) and (\ref{eq:Krauss-amp-damping}), we obtain
the average fidelity of the quantum state reconstructed by the receivers
${\rm Bob_{2}}$ and ${\rm Bob_{3}}$ as 
\begin{equation}
F_{AD}^{{\rm Bob}_{2}}=\frac{\left(\eta_{A}^{2}-\eta_{A}+2\right)\sin^{4}(\theta)+\left(\eta_{A}^{2}-\eta_{A}+2\right)\cos^{4}(\theta)-2\sin^{2}(\theta)\cos^{2}(\theta)\left(\eta_{A}^{3}+(\eta_{A}-1)\eta_{A}^{2}\cos(2\phi)-2\eta_{A}^{2}+\eta_{A}-2\right)}{2\left(\eta_{A}^{2}+1\right)},\label{eq:AD-feditilty1}
\end{equation}
and
\begin{equation}
F_{AD}^{{\rm Bob}_{3}}=\frac{\sin^{2}(\theta)\cos^{2}(\theta)\left(-\sqrt{1-\eta_{A}}\eta_{A}^{2}+(1-\eta_{A})^{3/2}\eta_{A}\cos(2\phi)-\sqrt{1-\eta_{A}}\eta_{A}+2\eta_{A}+2\sqrt{1-\eta_{A}}\right)+\sin^{4}(\theta)+\cos^{4}(\theta)}{\eta_{A}+1},\label{eq:AD-feditilty2}
\end{equation}
respectively. It would be relevant to mention that the fidelity expressions
of the quantum state obtained by other lower power agent ${\rm Bob}_{1}$
are exactly the same as that for ${\rm Bob}_{3}$ in all the noise
models discussed in this paper. Hence, the state reconstructed by
both the lower power agents will be equally affected by the noise.
Henceforward, we will only report the fidelity expressions for ${\rm Bob}_{3}$
and the conclusions deduced from them will be automatically implementable
for ${\rm Bob}_{1}$ as well. Further, it would be appropriate to
note that the fidelity expressions reported in Eqs. (\ref{eq:AD-feditilty1})
and (\ref{eq:AD-feditilty2}) are averages of fidelity obtained for
all the possible cases, i.e., different measurement outcomes of both
the senders and the receivers whose cooperation is essential for the
reconstruction of the quantum state at the receiver's end. In particular,
when the higher (lower) power agent\textcolor{magenta}{{} }reconstructs
a quantum state, we need to compute fidelity expressions for 8 (16)
possible combinations of measurement outcomes of all the senders and
the other collaborating receivers. In all the following expressions
of fidelity and the figures shown here we are explicitly calculating
this average fidelity. 

\begin{figure}[H]
\begin{centering}
\includegraphics[scale=1.0,angle=270]{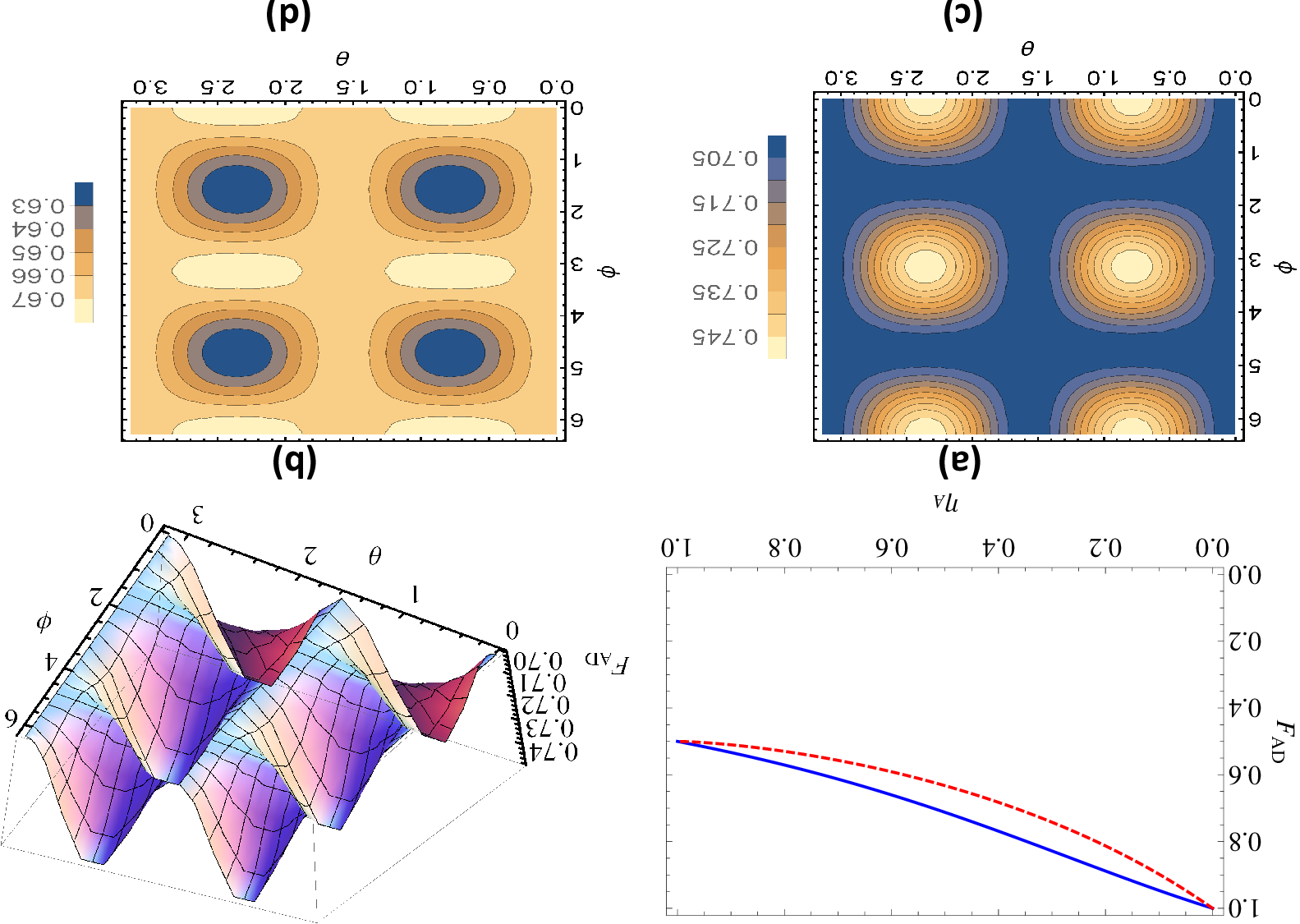}
\par\end{centering}

\protect\caption{\label{fig:AD1}(Color online) In (a), the fidelity $F_{AD}$ is plotted
with the decoherence rate $\eta_{A}$ considering the state parameters
$\theta=\frac{\pi}{4}$ and $\phi=\frac{\pi}{3}$. The smooth (blue)
and dotted (red) lines correspond to the fidelities of the quantum states for the higher
power agent ${\rm Bob_{2}}$ and the lower power agent ${\rm Bob_{3}}$,
respectively. (b) shows three dimensional variation of the fidelity
$F_{AD}^{{\rm Bob}_{2}}$ with quantum state parameters, i.e., amplitude
$(\theta)$ and phase $(\phi)$ for decoherence rate $\eta_{A}=\frac{1}{2}$.
(c) and (d) show the variation of the average fidelity of the quantum
state obtained by the receivers ${\rm Bob_{2}}$ and ${\rm Bob_{3}}$,
respectively, with the state parameters considering $\eta_{A}=\frac{1}{2}$. }
\end{figure}

The average fidelity ($F_{AD}$) depends on various parameters, such
as amplitude $(\theta)$ and phase $(\phi)$ of the quantum state
to be remotely prepared and the decoherence rate $(\eta_{A})$. This
dependence is illustrated in Fig. \ref{fig:AD1}. Specifically, Fig.
\ref{fig:AD1} a establishes that the state reconstructed by the higher
power agent ${\rm Bob_{2}}$ gets less affected due to noise in comparison
to the state reconstructed by the lower power agents. Fig. \ref{fig:AD1}
b illustrates\textcolor{magenta}{{} } variation of the fidelity with
state parameters for the higher power agent considering the decoherence
rate $\eta_{A}=\frac{1}{2}$. It can be inferred from the plot that
the obtained fidelity may be higher for certain choices of quantum
state to be remotely prepared. These facts can also be illustrated
using contour plots. Hereafter, we would only use contour plots to
investigate the effect of noise on the deterministic HJRSP scheme.
However, here we have shown a corresponding contour plot as well in
Fig. \ref{fig:AD1} c. A slightly different nature of dependence for
the average fidelity of the lower power agent on state parameters
can be observed in Fig. \ref{fig:AD1} d. 

The case reported here for a dissipative interaction with a vacuum
bath, simulated as AD channel, can be extended to generalized amplitude
damping \cite{subh,kishoe-qdistri} and squeezed generalized amplitude
damping \cite{subh,kishoe-qdistri} noise models, where finite temperature
reservoir are considered with zero and non-zero squeezing, respectively.
The same will be investigated separately.

\subsection{Phase damping (PD) noise:}

The PD noise model is represented by the following Kraus operators
\cite{nielsen,RSP-with-noise,preskill}

\begin{equation}
E_{0}^{PD}=\sqrt{1-\eta_{P}}\left[\begin{array}{cc}
1 & 0\\
0 & 1
\end{array}\right],\,\,\,\,\,\,\,\,\,\,\,\,\,\,\,E_{1}^{PD}=\sqrt{\eta_{P}}\left[\begin{array}{cc}
1 & 0\\
0 & 0
\end{array}\right],\,\,\,\,\,\,\,\,\,\,\,\,\,\,\,E_{2}^{PD}=\sqrt{\eta_{P}}\left[\begin{array}{cc}
0 & 0\\
0 & 1
\end{array}\right],\label{eq:Krauss-phase-damping}
\end{equation}
where $\eta_{P}$ ($0\leq\eta_{P}\leq1$) is the decoherence rate.
The PD noise simulates an interaction with the surroundings\textcolor{magenta}{{}
}when energy loss is not involved. In presence of this noise, the
average fidelities of the quantum state reconstructed by the higher
and lower power agents ${\rm Bob_{2}}$ and ${\rm Bob_{3}}$ can be
obtained using\textcolor{red}{{} }Eqs. (\ref{eq:fdly}), (\ref{eq:After noise})
and (\ref{eq:Krauss-phase-damping}) as 
\[
\begin{array}{lcl}
F_{PD}^{{\rm Bob}_{2}} & = & \frac{1}{4}\left(\eta_{P}^{2}-(\eta_{P}-2)\eta_{P}\cos(4\theta)-2\eta_{P}+4\right),\end{array}
\]
and
\begin{equation}
\begin{array}{lcl}
F_{PD}^{{\rm Bob}_{3}} & = & \frac{1}{4}\left(-\eta_{P}^{3}+\left(\eta_{P}^{2}-3\eta_{P}+3\right)\eta_{P}\cos(4\theta)+3\eta_{P}^{2}-3\eta_{P}+4\right),\end{array}\label{eq:PD-fidelity}
\end{equation}
respectively.

\begin{figure}[H]
\begin{centering}
\includegraphics[scale=0.7,angle=270]{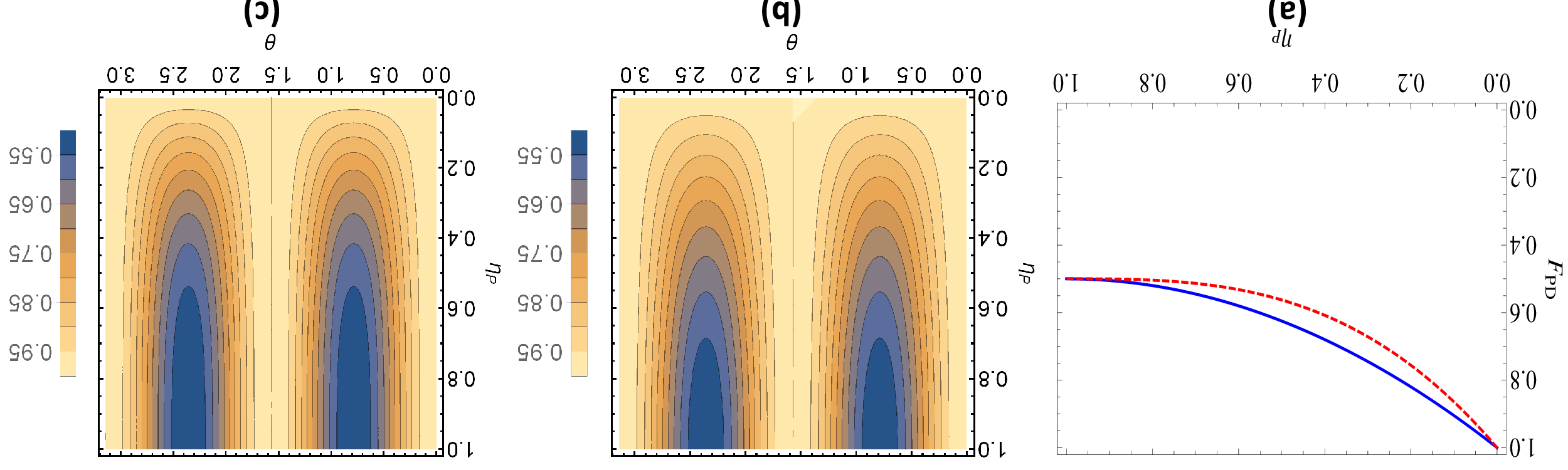}
\par\end{centering}

\protect\caption{\label{fig:PD}(Color online) In (a), the average fidelity $F_{PD}$
of the quantum state reconstructed by both lower and higher power
agents are plotted together with the decoherence rate $\eta_{P}$
in the presence of PD noise for $\theta=\frac{\pi}{4}$. The smooth
(blue) and dotted (red) lines correspond to the fidelity for the state reconstructed by higher
power agent and the lower power agent, respectively. (b) and (c) show
the variation of average fidelity with state parameter $\theta$ and
decoherence rate $\eta_{P}$ in contour plots for receivers ${\rm Bob_{2}}$
and ${\rm Bob_{3}}$, respectively. It can be observed that the contour
plot for fidelity variation of ${\rm Bob_{2}}$'s quantum state is symmetric with
that of ${\rm Bob_{3}}$. However, the higher power agent obtains
the state with slightly more fidelity than that of the lower power
agent. }
\end{figure}

It can be observed from the expressions of average fidelity in Eq.
(\ref{eq:PD-fidelity}) that it is phase $(\phi)$ independent. Consequently,
a family of states with the same value of $\theta$ undergo the same
decoherence, i.e., for a particular value of $\theta$ all the states
will be affected in the same way for any value of $\phi$. Here, in
the HJRSP scheme, this corresponds to the fact that any choice of
${\rm Alice_{2}}$ will not affect the state reconstructed by the
receivers. Further, we have illustrated the dependence of average
fidelity on various parameters in Fig. \ref{fig:PD}. Specifically,
Fig. \ref{fig:PD} a shows a similar nature of what was observed in
the presence of AD noise in Fig. \ref{fig:AD1} a, i.e., if the higher
power agent ${\rm Bob_{2}}$ reconstructs the state, then the reconstructed
state gets less affected by noise in comparison to the cases where
the lower power agents reconstruct the state. In Fig. \ref{fig:PD}
b and c, the contour plots of variation of the average fidelity with
the amplitude $(\theta)$ and decoherence rate $(\eta_{P})$ are shown
for ${\rm Bob_{2}}$ and ${\rm Bob_{3}}$, respectively. The contour
plots\textcolor{magenta}{{} }show that although a symmetric variation
in the fidelity is observed, the state reconstructed by the higher
power agent is found to be less affected by the noise. Further, it
is observed that\textbf{ }for certain values of different parameters
we may obtain a unit fidelity even in the noisy environment.

\subsection{Collective dephasing (CD) noise:}

The CD noise model is represented by the following unitary (phase)
operator 
\begin{equation}
U_{CD}=\left[\begin{array}{cc}
1 & 0\\
0 & \exp(i\Phi)
\end{array}\right],\label{eq:Collective-dephasing-noise}
\end{equation}
where $\Phi$ is the noise parameter that may change with time, but
remains the same at an instant for all the qubits traveling simultaneously
through the noisy channel. It should be noted here that the collective
noise models consider a coherent effect of interaction on all the
travel qubits. The average fidelities for the state reconstructed
by ${\rm Bob_{2}}$ and ${\rm Bob_{3}}$ can be obtained under CD
noise model as 
\begin{equation}
\begin{array}{lcl}
F_{CD}^{{\rm Bob_{2}}} & = & \frac{1}{16}(-\cos(4\theta-2\Phi)-\cos(4\theta+2\Phi)+2\cos(4\theta)+2\cos(2\Phi)+14),\\
F_{CD}^{{\rm Bob_{3}}} & = & \sin^{2}(\theta)\cos^{2}(\theta)(\cos(\Phi)+\cos(3\Phi))+\sin^{4}(\theta)+\cos^{4}(\theta),
\end{array}\label{eq:CD-fidelity}
\end{equation}
respectively.

\begin{figure}[H]
\begin{centering}
\includegraphics[scale=0.7,angle=270]{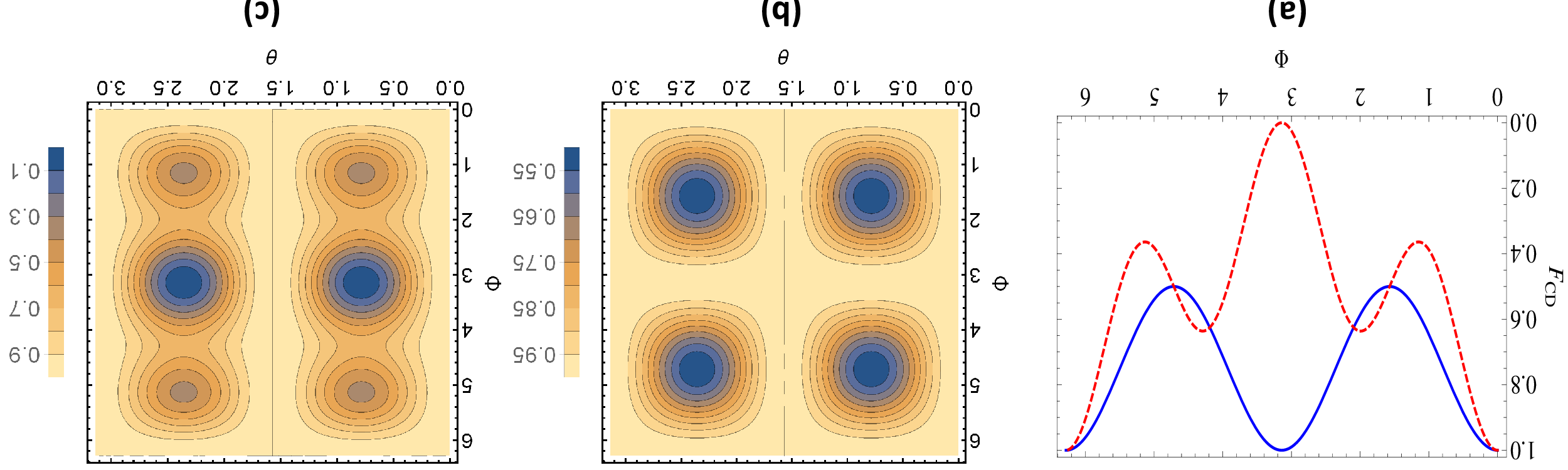}
\par\end{centering}

\protect\caption{\label{fig:CD}(Color online) In (a), the average fidelity $F_{CD}$
is shown to vary with the noise parameter ($\Phi$) of CD noise for
$\theta=\frac{\pi}{4}$. The smooth (blue) and dotted (red) lines
correspond to the fidelities of the quantum states obtained by the higher and lower power agents,
respectively. (b) and (c) show the variation of the average fidelity
with a noise parameter and the amplitude of the state to be remotely
prepared for the receivers ${\rm Bob_{2}}$ and ${\rm Bob_{3}}$,
respectively. }
\end{figure}

From the above expressions, it can be observed that the average fidelity
is free from the phase of the quantum state to be remotely prepared.
Hence, the fidelity of the obtained state will only depend on the
noise parameter and the amplitude of the state to be reconstructed.
To elaborate the dependence explicitly, we have plotted the average
fidelity ($F_{CD}$) as a function of amplitude $(\theta)$ and the
noise parameter $(\Phi)$ in Fig. \ref{fig:CD}. Specifically, in
Fig. \ref{fig:CD} a, the average fidelity is shown to vary with the
noise parameter for a family of states with an arbitrary value of
phase angle and amplitude $\theta=\frac{\pi}{4}$. Here, it is interesting
to note that, unlike the previous cases when the HJRSP scheme was
subjected to AD and PD noise models, even the\textcolor{magenta}{{}
}lower \textcolor{magenta}{{} }power agent can acquire the quantum state
with higher fidelity than that of the higher power agent. Specifically,
in Fig. \ref{fig:CD} a, in the vicinity of $\Phi=\frac{\pi}{2}$
the state reconstructed by a lower power agent is found to be less
affected by noise in comparison to the same state reconstructed by
a higher power agent. Further, when the noise parameter $\Phi=\pi$
then ${\rm Bob_{2}}$ recovers unaffected state, i.e., the state with
unit fidelity, while ${\rm Bob_{3}}$ can reconstruct the quantum
state with negligible fidelity. Fig. \ref{fig:CD} b and c further
illustrate the same facts through contour plots and manifests the
effect of noise on all the possible quantum states that can be remotely
prepared. It can be observed from the plots that unit fidelity can
be obtained in some cases. Except a few values of the parameters,
in general, the higher power agent can extract higher fidelity quantum
state.

\subsection{Collective rotation (CR) noise:}

The CR noise model is represented by a unitary rotation operator 
\begin{equation}
U_{CR}=\left[\begin{array}{cc}
\cos\Theta & -\sin\Theta\\
\sin\Theta & \cos\Theta
\end{array}\right],\label{eq:Collective-rotation-noise}
\end{equation}
where $\Theta$ is the noise parameter and have the same effect and
property as $\Phi$ in the CD noise model. The expressions for the
average fidelities of the quantum state reconstructed by ${\rm Bob_{2}}$
and ${\rm Bob_{3}}$ are a bit complex and to ensure that the flow
of the paper is not disturbed, they are reported in Appendix A in
Eq. (\ref{eq:CR-fidelity}).

\begin{figure}[H]
\begin{centering}
\includegraphics[scale=0.7,angle=270]{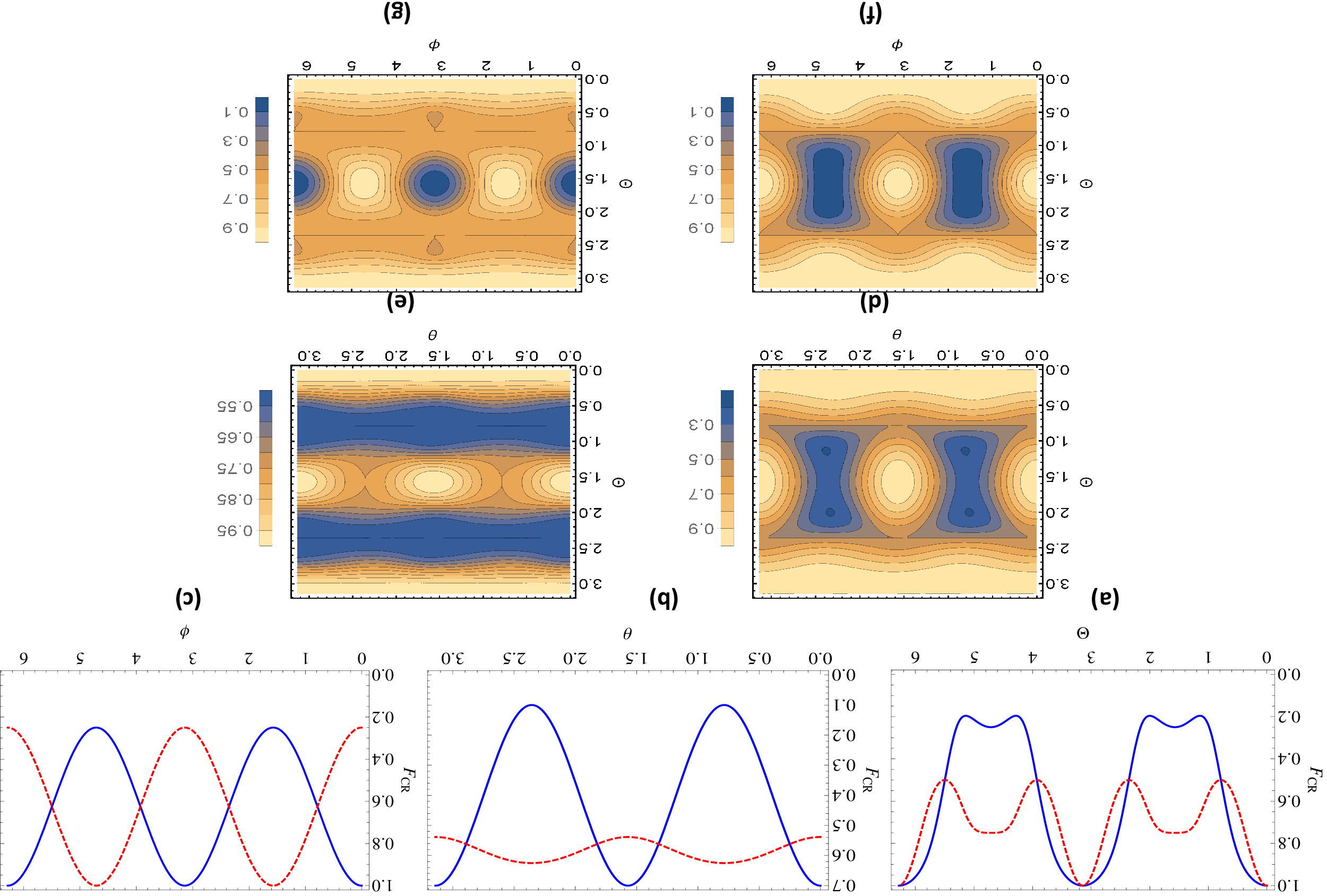}
\par\end{centering}

\protect\caption{\label{fig:CR}(Color online) In (a), the average fidelity $F_{CR}$
is plotted together for both higher and lower power agents with the
noise parameter $\Theta$ for state parameters $\theta=\pi/4$ and
$\phi=\pi/3$. (b) illustrates the dependence of the average fidelity
$F_{CR}$ of the reconstructed state for the agents on the amplitude
parameter $\theta$ for $\phi=\pi/2$ and noise parameter $\Theta=\pi/3$.
In (c), the average fidelity is plotted with the phase angle $\phi$
for $\theta=\pi/3$ and $\Theta=\pi/2$. In (a)-(c) the smooth (blue)
and dotted (red) lines correspond to the fidelities of state reconstructed
by higher power agent ${\rm Bob_{2}}$ and the lower power agent ${\rm Bob_{3}}$,
respectively. (d) and (e) show the variation of average fidelity with
noise parameter and amplitude parameter considering $\phi=\frac{\pi}{3}$
for the receivers ${\rm Bob_{2}}$ and ${\rm Bob_{3}}$, respectively.
Similarly, (f) and (g) demonstrate the variation for $\theta=\frac{\pi}{4}$
with noise parameter and phase angle for the higher and lower power
agents, respectively.}
\end{figure}

To express the dependence of average fidelity of the state received
by both higher and lower power agents on various parameters we have
plotted the fidelity expressions in Fig. \ref{fig:CR}. Specifically,
Fig. \ref{fig:CR} a shows the variation in the average fidelity $F_{CR}$
with the noise parameter $(\Theta)$ for a particular quantum state
with $\theta=\pi/4$ and $\phi=\pi/3$ to be remotely prepared. Interestingly,
for a few specific values of the noise parameter (e.g., for $\frac{\pi}{3}<\Theta<\frac{2\pi}{3}$),
the higher power agent can reconstruct the state with lesser fidelity
compared to the state obtained by the lower power agent. In contrast
to the CD noise case, here lower power agent can also reconstruct
a state with unit fidelity. For a particular value of noise parameter,
i.e., $\Theta=\pi/3$, the class of states with phase angle $\frac{\pi}{2}$
show that they are not equally affected due to noise (cf. Fig. \ref{fig:CR}
b). In fact, if the state is reconstructed by\textcolor{magenta}{{}
}the higher power agent then the value of the amplitude parameter
matters more than the cases when lower power agents choose to do so.
Further, it also shows that for $\theta=\pi/2$, higher power agent
${\rm Bob_{2}}$ achieves the maximum fidelity of the received state,
whereas the fidelity of the state obtained by the lower power agent
${\rm Bob_{3}}$ could never reach that limit. Fig. \ref{fig:CR}
c shows the fidelity variation with phase angle $\phi$, and a complimentary
nature for the fidelity of the quantum state obtained by higher and
lower power agents is observed, i.e., it is found that if the fidelity
for one increases, it decreases for the other. The contour plots in
Fig. \ref{fig:CR} d-g show that some specific choice of initial quantum
states can be remotely prepared with unit fidelity or fidelity close
to 1 in the CR noisy environment for particular values of the noise
parameter.

The nature observed here for the collective noise model, i.e., the
effect of collective noise is in contrast of that in the AD and PD
channels, is consistent with a few earlier observations \cite{Kishore-decoy,vishal}.
Specifically, in Ref. \cite{Kishore-decoy}, it has been observed
that single qubits perform better as decoy qubits in secure quantum
communication than entangled qubits when subjected to AD and PD noise;
whereas completely opposite results have been obtained in collective
noise, i.e., a few Bell states were shown to be decoherence free.
Similarly, single-qubit-based secure quantum communication schemes
have been found robust in AD and PD channels while their entangled-state-based
counterparts were found to perform better in the presence of collective
noise \cite{vishal}.

\subsection{Pauli (P) noise:}

There are four well known Pauli matrices that are frequently used
in quantum information. Just to introduce the notation followed in
this section, we note that the Pauli matrices are described in the
following manner: 
\begin{equation}
\sigma_{1}\equiv\sigma_{x}\equiv\left[\begin{array}{cc}
0 & 1\\
1 & 0
\end{array}\right],\,\,\,\,\,\,\,\,\sigma_{2}\equiv\sigma_{y}\equiv\left[\begin{array}{cc}
0 & -i\\
i & 0
\end{array}\right],\,\,\,\,\,\,\,\,\sigma_{3}\equiv\sigma_{z}\equiv\left[\begin{array}{cc}
1 & 0\\
0 & -1
\end{array}\right],\,\,\,\,\,\,\sigma_{4}\equiv I\equiv\left[\begin{array}{cc}
1 & 0\\
0 & 1
\end{array}\right].\label{eq:Pauli-matrices}
\end{equation}
These operators are relevant for the present study as a quantum state
described by the density operator $\rho$, evolves under Pauli noise
model (in Pauli channel) as described in Eq. (\ref{eq:After noise}).
Specifically, in Pauli channels, a single qubit state will evolve
as 
\begin{equation}
\rho^{\prime}=\sum_{i=1}^{4}E_{i}^{P}\rho E_{i}^{P\dagger},\label{eq:pauli-noise}
\end{equation}
where $E_{i}^{P}=\sqrt{p_{i}}\sigma_{i}$: $p_{i}$ is the probability
of $i$th type of error, which is modeled by Pauli operator (error
operator) $\sigma_{i}$ defined in Eq. (\ref{eq:Pauli-matrices}),
and $\sum_{i}^{4}p_{i}=1$. To visualize that $p_{i}$ is the probability
of the error modeled by $\sigma_{i}$, we may expand Eq. (\ref{eq:pauli-noise})
as 
\begin{equation}
\begin{array}{lcc}
\rho^{\prime} & = & p_{1}\sigma_{1}\rho\sigma_{1}+p_{2}\sigma_{2}\rho\sigma_{2}+p_{3}\sigma_{3}\rho\sigma_{3}+(1-p_{1}-p_{2}-p_{3})\rho,\end{array}\label{eq:pauli-noise-model}
\end{equation}
which clearly shows that the density operator $\rho$ undergoes a
bit-flip (modeled by $\sigma_{1})$ with the probability $p_{1}$,
a combined bit and phase-flip (modeled by $\sigma_{2})$ with probability
$p_{2}$ and a phase flip (modeled by $\sigma_{3})$ with the probability
$p_{3}$. However, $\rho$ remains unchanged with the probability
$(1-p_{1}-p_{2}-p_{3})=p_{4}$, which implies an error free channel.
The Pauli channel is widely discussed in the literature \cite{Pauli-exp,Pauli-ch-est,Pauli-ch-cap},
and its name originates from the error operators of the channel. Specifically,
the most general Pauli channel described by (\ref{eq:pauli-noise-model})
reduces to (i) a bit flip channel when $p_{2}=p_{3}=0,$ (ii) a bit-phase
flip channel when $p_{1}=p_{3}=0,$ (iii) a phase flip channel when
$p_{1}=p_{2}=0,$ and (iv) a noise free channel or identity channel
when $p_{1}=p_{2}=p_{3}=0.$ Similarly, we may define a depolarizing
noise channel when all the bit-flip, phase-flip, and bit-phase-flip
errors occur with equal probability $p_{1}=p_{2}=p_{3}=p\leq\frac{1}{3}$,
and the state remains unchanged with the remaining probability.

Above discussion shows that if we can obtain fidelity for the most
general Pauli channel described by (\ref{eq:pauli-noise-model}),
we can obtain the fidelities for evolution under specific noise models
as special cases of that. Keeping this in mind, we have obtained following
analytic expressions for the average fidelity of the quantum state
remotely prepared by the proposed HJRSP scheme when the quantum state
to be prepared remotely is evolved under most general Pauli noise
model, and either ${\rm Bob_{2}}$ or ${\rm Bob_{3}}$ reconstructs
the state: 
\begin{equation}
\begin{array}{lcl}
F_{P}^{{\rm Bob_{2}}} & = & \frac{1}{(p_{1}+p_{2})^{2}+(p_{3}+p_{4})^{2}}((p{}_{1}^{2}+p{}_{2}^{2}+p_{1}(2p_{2}-p_{3}-p_{4})-p_{2}(p_{3}+p_{4})+(p_{3}+p_{4})^{2})\cos^{4}(\theta)\\
 & + & 2\cos^{2}(\theta)(p{}_{3}^{4}-p{}_{1}^{3}(p_{3}-3p_{4})+p{}_{2}^{3}(3p_{3}-p_{4})+p{}_{4}^{4}+p_{2}(p_{3}+p_{4})^{3}+3p{}_{2}^{2}(p{}_{3}^{2}+p{}_{4}^{2})\\
 & + & p_{1}(p{}_{2}^{2}p_{4}+(p_{3}+p_{4})^{3})+p{}_{1}^{2}(p_{2}p_{3}+3(p{}_{3}^{2}+p{}_{4}^{2}))+(p{}_{1}^{4}+p{}_{1}^{2}(-2p{}_{2}^{2}+(p_{3}-p_{4})^{2})\\
 & - & 2p_{1}(p_{3}-p_{4})(p_{2}(-p_{3}+p_{4})+(p_{3}+p_{4})^{2})+p_{2}(p{}_{2}^{3}+p_{2}(p_{3}-p_{4})^{2}+2(p_{3}-p_{4})\\
 & \times & (p_{3}+p_{4})^{2}))\cos(2\phi))\sin^{2}(\theta)+(p{}_{1}^{2}+p{}_{2}^{2}+p_{1}(2p_{2}-p_{3}-p_{4})-p_{2}(p_{3}+p_{4})\\
 & + & (p_{3}+p_{4})^{2})\sin^{4}(\theta)+\frac{1}{2}(p{}_{1}^{2}(5p_{2}+6p_{3})p_{4}-2p_{3}p_{4}(-3p{}_{2}^{2}+p_{3}p_{4})+p_{1}p_{2}(5p_{2}p_{3}\\
 & + & 2(p_{3}+p_{4})^{2}))\sin^{2}(2\theta),\\
F_{P}^{{\rm Bob_{3}}} & = & (((p_{1}+p_{2})^{2}+(p_{3}+p_{4})^{2})\cos^{4}(\theta)+2\cos^{2}(\theta)(p_{3}(p{}_{1}^{3}-(p_{2}-p_{3})^{2}(p_{2}+p_{3})\\
 & + & p_{1}(11p{}_{2}^{2}+3p{}_{3}^{2}))+p_{2}(p{}_{1}^{2}+3p{}_{2}^{2})p_{4}+3p{}_{2}^{2}p{}_{4}^{2}+(3p_{1}+p_{2})p{}_{4}^{3}+p{}_{4}^{4}+(p_{1}-p_{2}\\
 & - & p_{3}+p_{4})(p{}_{1}^{3}-p{}_{1}^{2}p_{2}+p_{2}(p{}_{2}^{2}+(p_{3}-p_{4})(3p_{3}+p_{4}))-p_{1}(p{}_{2}^{2}+(p_{3}-p_{4})(p_{3}+3p_{4})))\\
 & \times & \cos(2\phi))\sin^{2}(\theta)+((p_{1}+p_{2})^{2}+(p_{3}+p_{4})^{2})\sin^{4}(\theta)+\frac{1}{2}(5p{}_{1}^{3}p_{4}+p{}_{1}^{2}(5p_{3}(p_{2}+p_{3})\\
 & + & 4p_{3}p_{4}+7p{}_{4}^{2})+p_{3}p_{4}(12p{}_{2}^{2}+7p_{2}(p_{3}+p_{4})+2(p_{3}-p_{4})(p_{3}+p_{4}))+p1(7p{}_{2}^{2}p_{4}\\
 & + & 5p_{3}p_{4}(p_{3}+p_{4})+2p_{2}(p_{3}+p_{4})(5p_{3}+3p_{4})))\sin^{2}(2\theta)).
\end{array}\label{eq:Pauli-fidelity}
\end{equation}

\begin{figure}[H]
\begin{centering}
\includegraphics[scale=1.0,angle=270]{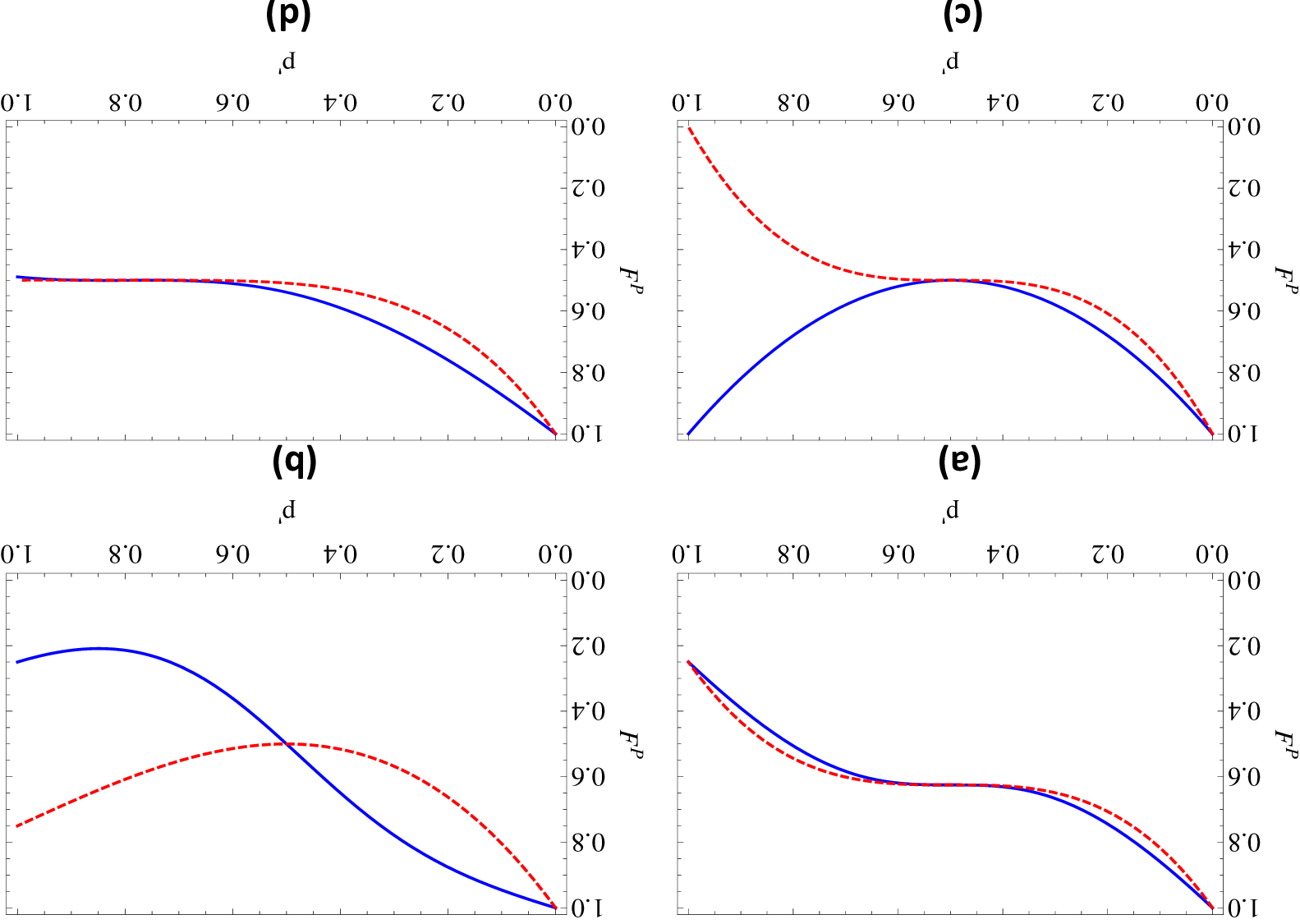}
\par\end{centering}

\protect\caption{\label{fig:Pauli}(Color online) The average fidelity of the quantum
state reconstructed by both the higher and lower power agents when
subjected to Pauli noise are shown. (a), (b), (c), and (d) correspond
to bit-flip, bit-phase-flip, phase-flip and depolarizing channels,
respectively. In all these plots, $\theta=\pi/4$ and $\phi=\pi/3$.
The smooth (blue) and dotted (red) lines correspond to the fidelities
of the quantum states obtained by the higher power agent and the lower power agent, respectively. }
\end{figure}

To illustrate the effect of Pauli noise, we have plotted the average
fidelity ($F_{P}$) under the Pauli noise model for the state parameters
$\theta=\pi/4$ and $\phi=\pi/3$ in Fig. \ref{fig:Pauli}. Here,
we have used a depolarizing channel, where the total probability of
error for bit, phase and bit-phase flip errors is $p^{\prime}$, and
consequently, $\frac{p^{\prime}}{3}$ is the\textcolor{magenta}{{} }probability
for each kind of error. Specifically, in Fig. \ref{fig:Pauli} (a),
the average fidelity $F_{P}$ is shown in case of bit-flip errors.
It can be observed that the fidelities of the state obtained by ${\rm Bob_{2}}$
and ${\rm Bob_{3}}$ vary almost similarly and decay to the lowest
value when the probability of errors $p'$ is unity. Similarly, in
Fig. \ref{fig:Pauli} (b), the average fidelity of the state reconstructed
by both higher and lower power agents in bit-phase-flip channel is
shown. Interestingly, it has been observed that for higher probability
of errors, i.e., $p'>0.5$, even higher power agent ${\rm Bob_{2}}$
has less fidelity of the reconstructed state than that of the lower
power agent. In fact, beyond this value of probability of errors a
revival in the fidelity for the lower power agent can be seen. In
Fig. \ref{fig:Pauli} (c), variation of the average fidelity $F_{P}$
with the probability of phase-flip error is illustrated. In contrast
with Fig. \ref{fig:Pauli} (b), here the fidelity obtained by the
higher power agent is revived and reached to
unity for maximum probability of error. Further, the fidelity of the
state obtained by the higher power agent is always
higher than that of the lower power agent, which shows a gradual decay
with increasing probability of error (cf. red (dashed) line in Fig. \ref{fig:Pauli} (c)). Finally, in Fig. \ref{fig:Pauli}
(d), we have plotted the average fidelity $F_{P}$ under the effect
of depolarizing channel, where the fidelity of states obtained by
both the receivers converges for $0.6<p'<0.9$, and after that it
decays for the higher power agent.

\section{Conclusions\label{sec:Conclusions} }

In the present paper we have introduced the notion of HJRSP and have proposed a protocol for deterministic
HJRSP using 5-qubit cluster state of the form (\ref{eq:state}). The
scheme is also illustrated through Fig. \ref{fig:HJRSP-scheme}. It is
unique in nature because it allows joint preparability (more than
one sender), a feature which is desirable in practical\textcolor{magenta}{{}
}situations \textcolor{magenta}{{} }as illustrated in Sec. \ref{sec:Practical-Applications},
but not present in any of the existing schemes of hierarchical quantum
communication. Further, in the present work, the effect of various
noise models on the proposed HJRSP scheme has been investigated in
detail. In contrast, in none of the existing proposals for hierarchical
quantum communication, the effect of noise has been discussed. However,
in a practical situation, it is impossible to circumvent the existence
of noise. Thus, the present study not only makes it more realistic,
it also yields an extremely relevant and essential feature of hierarchical
quantum communication. Further, in addition to the deterministic HJRSP,
a protocol for probabilistic HJRSP using a non-maximally entangled
cluster state is also introduced here. Usually, a reduction of required quantum
resources is expected in probabilistic RSP. However, it was not obtained
here due to use of a non-maximally entangled state. 

The study of the effect of different noise models has led to many
interesting results. Specifically, it is observed that the quantum
state that the higher power agent and the lower power agent can reconstruct
may get decohered due to interaction with its surroundings, which have
been analyzed and discussed in detail in the previous section. In
brief, the higher power agent can reconstruct the quantum state less
affected due to AD and PD noise than that of the lower power agent.
However, when the travel qubits are subjected to collective or Pauli
noise, lower power agent may also get better performance than that
of the higher power agent. Interestingly, a few specific quantum states
can be remotely prepared in an unaffected manner, even in the presence of noise. Here,
we have restricted ourselves to Markovian noise. The effect of non-Markovian
type of noise will be reported elsewhere.

It would be interesting to study controlled versions of the hierarchical
quantum communication protocols proposed here and in the recent past,
as a variety of applications of the various controlled quantum communication
protocols have been widely discussed \cite{crypt-switch}. We expect that
due to the wide applicability of the hierarchical schemes the proposed
scheme may be of interest to experimentalists and in the near future
the results reported here will be experimentally verified.

\textbf{Acknowledgment:} CS thanks Japan Society for the Promotion
of Science (JSPS), Grant-in-Aid for JSPS Fellows no. 15F15015. AP
thanks Department of Science and Technology (DST), India for the support
provided through the project number EMR/2015/000393. Authors thank M. Ozawa for his interest in the work, and for some useful suggestions and comments. Authors thank Roopal Vegad for her help in preparing a schematic diagram.

\section*{Appendix A}

The average fidelity expressions of the quantum states reconstructed by the higher and lower power agents
when the HJRSP scheme is subjected to CR noise are obtained as

\[
\begin{array}{lcl}
F_{CR}^{{\rm Bob_{2}}} & = & \frac{1}{8((3+\cos(4\Theta))^{2}-4\cos^{2}(\Theta)(\cos(4\Theta-2\theta)+3\cos(2\theta))^{2}\cos^{2}(\phi)\sin^{2}(\Theta))}\left\{ (3+\cos^{2}(4\Theta)(5+3\cos(4\Theta))\cos^{4}(\Theta)\right.\\
 & - & (42+88\cos(2\Theta)+36\cos(4\Theta)+23\cos(6\Theta)+2\cos(8\Theta)+\cos(10\Theta)+9\cos(6\Theta-4\theta)\\
 & + & 4\cos(8\Theta-4\theta)+2\cos(10\Theta-4\theta)+7\cos(2(\Theta-2\theta))+18\cos(4(\Theta-\theta))+24\cos(4\theta)\\
 & + & 2\cos(4(\Theta+\theta))-\cos(2(\Theta+2\theta))-\cos(6\Theta+4\theta))\cos(2\phi)\sin^{2}(\Theta)-16(5\cos(2\Theta)+\\
 & + & \cos(6\Theta))\cos^{3}(\theta)\sin^{3}(2\Theta)\sin(\theta)+2\cos^{2}(\Theta)(78+14\cos(2\Theta)+40\cos(4\Theta)-9\cos(6\Theta)\\
 & + & 10\cos(8\Theta)-5\cos(10\Theta))\cos^{2}(\theta)\sin^{2}(\theta)+16(5\cos(2\Theta)+\cos(6\Theta))\cos(\theta)\sin^{3}(2\Theta)\sin^{3}(\theta)\\
 & + & \left.(3+\cos^{2}(4\Theta))(5+3\cos(4\Theta))\sin^{4}(\theta)\right\} ,
\end{array}
\]
and

\begin{equation}
\begin{array}{lcl}
F_{CR}^{Bob_{3}} & = & \left[A+B+C+D+E+F+G+H\right],
\end{array}\label{eq:CR-fidelity}
\end{equation}
where
\[\begin{array}{lcl}
A & = & \left\{ 1/\left[(8-\cos(2\Theta-\phi)+\cos(6\Theta-\phi)+\cos(4\Theta-2\theta-\phi)-\cos(4\Theta+2\theta-\phi)-\cos(2\Theta+\phi)\right.\right.\\
 & + & \cos(6\Theta+\phi)+\cos(4\Theta-2\theta+\phi)-\cos(4\Theta+2\theta+\phi)+2\sin(4\Theta-2\theta)+2\sin(4\Theta+2\theta)\\
 & - & \left.\left.2\sin(2\Theta-2\theta-\phi)-2\sin(2\Theta+2\theta-\phi)-2\sin(2\Theta-2\theta+\phi)-2\sin(2\Theta+2\theta+\phi)\right]\right\} \\
 & \times & \{\cos^{8}(\Theta)+\cos^{4}(\theta)\sin^{8}(\Theta)-2\cos^{2}(\theta)\cos(2\phi)\sin^{8}(\Theta)\sin^{2}(\theta)+\sin^{8}(\Theta)\sin^{4}(\theta)\\
 & + & 2\cos^{4}(\Theta)\sin^{4}(\Theta)(\cos^{4}(\theta)+2\cos^{3}(\theta)(1-4\cos(\phi)+\cos(2\phi))\sin(\theta)+\cos^{2}(\theta)\cos(2\phi)\sin^{2}(\theta)\\
 & - & 4\cos(\theta)\cos^{2}(\phi)\sin^{3}(\theta)+\sin^{4}(\theta))+2\cos^{2}(\Theta)\sin^{6}(\Theta)(2\cos^{4}(\theta)\cos(\phi)-4\cos^{3}(\theta)\cos^{2}(\phi)\sin(\theta)\\
 & + & \cos^{2}(\theta)(3+\cos(2\phi))\sin^{2}(\theta)-4\cos(\theta)\cos(\phi)\sin^{3}(\theta)+4\cos^{2}(\frac{\phi}{2})\sin^{4}(\theta))+\cos(\Theta)\cos(2\theta)\sin^{7}(\Theta)\\
 & \times & (-2+2\cos(\phi)(-1+\sin(2\theta)))+\frac{3}{32}\sin^{4}(2\Theta)\sin^{2}(2\theta)+2\cos^{7}(\Theta)\sin(\Theta)(-\cos^{2}(\theta)(-1+\cos(\phi))\\
 & + & (-1+\cos(\phi))\sin^{2}(\theta)+\cos(\phi)\sin(2\theta))-4\cos^{6}(\Theta)\sin^{2}(\Theta)(\cos^{4}(\theta)(-1+\cos(\phi))+\cos(\phi)\sin^{4}(\theta)\\
 & + & 4\cos(\theta)\cos(\phi)\sin^{3}(\theta)\sin^{2}(\frac{\phi}{2})-\cos^{2}(\theta)\sin^{2}(\theta)\sin^{2}(\phi))+\cos^{3}(\Theta)\sin^{5}(\Theta)(-2\cos(2\theta)(1+\cos(\phi))\\
 & + & \cos(\phi)(2+2\cos(4\theta)-\cos(4\theta-\phi)+2\cos(\phi)-\cos(4\theta+\phi)-2\sin(2\theta)+\sin(4\theta-\phi)+\sin(4\theta+\phi)))\\
 & + & \cos^{5}(\Theta)\sin^{3}(\Theta)(-2\cos(2\theta)(-1+\cos(\phi))+\cos(\phi)(-2-2\cos(4\theta)+\cos(4\theta-\phi)-2\cos(\phi)\\
 & + & \cos(4\theta+\phi)-\sin(4\theta)+\sin(4\theta-\phi)+\sin(4\theta+\phi)))\},
\end{array}
\]
\[\begin{array}{lcl}
B & = & \left\{ 1/\left[8+\cos(2\Theta-\phi)-\cos(6\Theta-\phi)-\cos(4\Theta-2\theta-\phi)+\cos(4\Theta+2\theta-\phi)+\cos(2\Theta+\phi)\right.\right.\\
 & - & \cos(6\Theta+\phi)-\cos(4\Theta-2\theta+\phi)+\cos(4\Theta+2\theta+\phi)+2\sin(4\Theta-2\theta)+2\sin(4\Theta+2\theta)\\
 & + & \left.\left.2\sin(2\Theta-2\theta-\phi)+2\sin(2\Theta+2\theta-\phi)+2\sin(2\Theta-2\theta+\phi)+2\sin(2\Theta+2\theta+\phi)\right]\right\} \\
 & \times & \{\cos^{8}(\Theta)+\cos^{4}(\theta)\sin^{8}(\Theta)-2\cos^{2}(\theta)\cos(2\phi)\sin^{8}(\Theta)\sin^{2}(\theta)+\sin^{8}(\Theta)\sin^{4}(\theta)\\
 & + & 2\cos^{7}(\Theta)\sin(\Theta)(\cos^{2}(\theta)(1+\cos(\phi))-2\cos(\theta)\cos(\phi)\sin(\theta)-(1+\cos(\phi))\sin^{2}(\theta))\\
 & + & 2\cos^{4}(\Theta)\sin^{4}(\Theta)(\cos^{4}(\theta)+2\cos^{3}(\theta)(1+4\cos(\phi)+\cos(2\phi))\sin(\theta)+\cos^{2}(\theta)\cos(2\phi)\sin^{2}(\theta)\\
 & - & 4\cos(\theta)\cos^{2}(\phi)\sin^{3}(\theta)+\sin^{4}(\theta))+2\cos^{2}(\Theta)\sin^{6}(\Theta)(-2\cos^{4}(\theta)\cos(\phi)-4\cos^{3}(\theta)\cos^{2}(\phi)\sin(\theta)\\
 & + & \cos^{2}(\theta)(3+\cos(2\phi))\sin^{2}(\theta)+4\cos(\theta)\cos(\phi)\sin^{3}(\theta)-2(-1+\cos(\phi))\sin^{4}(\theta))-\cos(\Theta)\cos(2\theta)\sin^{7}(\Theta)\\
 & \times & (2+2\cos(\phi)(-1+\sin(2\theta)))+\frac{3}{32}\sin^{4}(2\Theta)\sin^{2}(2\theta)+4\cos^{6}(\Theta)\sin^{2}(\Theta)(\cos^{4}(\theta)(1+\cos(\phi))\\
 & + & 4\cos(\theta)\cos^{2}(\frac{\phi}{2})\cos(\phi)\sin^{3}(\theta)+\cos(\phi)\sin^{4}(\theta)+\cos^{2}(\theta)\sin^{2}(\theta)\sin^{2}(\phi))+\cos^{3}(\Theta)\sin^{5}(\Theta)(2\cos(2\theta)\\
 & \times & (-1+\cos(\phi))+\cos(\phi)(-2-2\cos(4\theta)-\cos(4\theta-\phi)+2\cos(\phi)-\cos(4\theta+\phi)+2\sin(2\theta)+\sin(4\theta-\phi)\\
 & + & \sin(4\theta+\phi)))+\cos^{5}(\Theta)\sin^{3}(\Theta)(2\cos(2\theta)(1+\cos(\phi))+\cos(\phi)(2+2\cos(4\theta)+\cos(4\theta-\phi)-2\cos(\phi)\\
 & + & \cos(4\theta+\phi)+\sin(4\theta)+\sin(4\theta-\phi)+\sin(4\theta+\phi)))\},\end{array}
\]
\[\begin{array}{lcl}
C & = & \left\{ 1/\left[-8-\cos(2\Theta-\phi)+\cos(6\Theta-\phi)-\cos(4\Theta-2\theta-\phi)+\cos(4\Theta+2\theta-\phi)-\cos(2\Theta+\phi)\right.\right.\\
 & - & \cos(6\Theta+\phi)+\cos(4\Theta-2\theta+\phi)+\cos(4\Theta+2\theta+\phi)+2\sin(4\Theta-2\theta)+2\sin(4\Theta+2\theta)\\
 & + & \left.\left.2\sin(2\Theta-2\theta-\phi)+2\sin(2\Theta+2\theta-\phi)+2\sin(2\Theta-2\theta+\phi)+2\sin(2\Theta+2\theta+\phi)\right]\right\} \\
 & \times & \{\cos^{8}(\Theta)+\cos^{4}(\theta)\sin^{8}(\Theta)-2\cos^{2}(\theta)\cos(2\phi)\sin^{8}(\Theta)\sin^{2}(\theta)+\sin^{8}(\Theta)\sin^{4}(\theta)\\
 & - & 2\cos^{7}(\Theta)\sin(\Theta)(\cos^{2}(\theta)(1+\cos(\phi))-2\cos(\theta)\cos(\phi)\sin(\theta)-(1+\cos(\phi))\sin^{2}(\theta))\\
 & + & 2\cos^{4}(\Theta)\sin^{4}(\Theta)(\cos^{4}(\theta)+4\cos^{3}(\theta)\cos^{2}(\phi)\sin(\theta)+\cos^{2}(\theta)\cos(2\phi)\sin^{2}(\theta)\\
 & - & 2\cos(\theta)(1+4\cos(\phi)+\cos(2\phi))\sin^{3}(\theta)+\sin^{4}(\theta))+2\cos^{2}(\Theta)\sin^{6}(\Theta)(-2\cos^{4}(\theta)(-1+\cos(\phi))\\
 & - & 4\cos^{3}(\theta)\cos(\phi)\sin(\theta)+\cos^{2}(\theta)(3+\cos(2\phi))\sin^{2}(\theta)+4\cos(\theta)\cos^{2}(\phi)\sin^{3}(\theta)-2\cos(\phi)\sin^{4}(\theta))\\
 & + & \frac{3}{32}\sin^{4}(2\Theta)\sin^{2}(2\theta)-2\cos(\Theta)\cos(2\theta)\sin^{7}(\Theta)(-1+\cos(\phi)(1+\sin(2\theta)))+4\cos^{6}(\Theta)\sin^{2}(\Theta)\\
 & \times & (\cos^{4}(\theta)\cos(\phi)-4\cos^{3}(\theta)\cos^{2}(\frac{\phi}{2})\cos(\phi)\sin(\theta)+2\cos^{2}(\frac{\phi}{2})\sin^{4}(\theta)+\cos^{2}(\theta)\sin^{2}(\theta)\sin^{2}(\phi))\\
 & - & \cos^{3}(\Theta)\sin^{5}(\Theta)(2\cos(2\theta)(-1+\cos(\phi))+\cos(\phi)(2+2\cos(4\theta)+\cos(4\theta-\phi)-2\cos(\phi)+\cos(4\theta+\phi)\\
 & + & 2\sin(2\theta)-\sin(4\theta-\phi)-\sin(4\theta+\phi)+\cos^{5}(\Theta)\sin^{3}(\Theta)(-2\cos(2\theta)(1+\cos(\phi))+\cos(\phi)(2+2\cos(4\theta)\\
 & + & \cos(4\theta-\phi)-2\cos(\phi)+\cos(4\theta+\phi)+\sin(4\theta)+\sin(4\theta-\phi)+\sin(4\theta+\phi)))\},
\end{array}
\]
\[\begin{array}{lcl}
D & = & \left\{ 1/\left[8-\cos(2\Theta-\phi)+\cos(6\Theta-\phi)-\cos(4\Theta-2\theta-\phi)+\cos(4\Theta+2\theta-\phi)-\cos(2\Theta+\phi)\right.\right.\\
 & + & \cos(6\Theta+\phi)-\cos(4\Theta-2\theta+\phi)+\cos(4\Theta+2\theta+\phi)-2\sin(4\Theta-2\theta)-2\sin(2(2\Theta+\theta))\\
 & + & \left.\left.2\sin(2\Theta-2\theta-\phi)+2\sin(2\Theta+2\theta-\phi)+2\sin(2\Theta-2\theta+\phi)+2\sin(2\Theta+2\theta+\phi)\right]\right\} \\
 & \times & \{\cos^{8}(\Theta)+\cos^{4}(\theta)\sin^{8}(\Theta)-2\cos^{2}(\theta)\cos(2\phi)\sin^{8}(\Theta)\sin^{2}(\theta)+\sin^{8}(\Theta)\sin^{4}(\theta)\\
 & + & 2\cos^{7}(\Theta)\sin(\Theta)(\cos^{2}(\theta)(-1+\cos(\phi))-2\cos(\theta)\cos(\phi)\sin(\theta)-(-1+\cos(\phi))\sin^{2}(\theta))\\
 & + & 2\cos^{4}(\Theta)\sin^{4}(\Theta)(\cos^{4}(\theta)+4\cos^{3}(\theta)\cos^{2}(\phi)\sin(\theta)+\cos^{2}(\theta)\cos(2\phi)\sin^{2}(\theta)\\
 & - & 2\cos(\theta)(1-4\cos(\phi)+\cos(2\phi))\sin^{3}(\theta)+\sin^{4}(\theta))+2\cos^{2}(\Theta)\sin^{6}(\Theta)(2\cos^{4}(\theta)(1+\cos(\phi))\\
 & + & 4\cos^{3}(\theta)\cos(\phi)\sin(\theta)+\cos^{2}(\theta)(3+\cos(2\phi))\sin^{2}(\theta)+4\cos(\theta)\cos^{2}(\phi)\sin^{3}(\theta)+2\cos(\phi)\sin^{4}(\theta))\\
 & + & \frac{3}{32}\sin^{4}(2\Theta)\sin^{2}(2\theta)+2\cos(\Theta)\cos(2\theta)\sin^{7}(\Theta)(1+\cos(\phi)(1+\sin(2\theta)))-4\cos^{6}(\Theta)\sin^{2}(\Theta)\\
 & \times & (\cos^{4}(\theta)\cos(\phi)+(-1+\cos(\phi))\sin^{4}(\theta)-4\cos^{3}(\theta)\cos(\phi)\sin(\theta)\sin^{2}(\frac{\phi}{2})-\cos^{2}(\theta)\sin^{2}(\theta)\sin^{2}(\phi))\\
 & + & \cos^{3}(\Theta)\sin^{5}(\Theta)(2\cos(2\theta)(1+\cos(\phi))+\cos(\phi)(2+2\cos(4\theta)-\cos(4\theta-\phi)+2\cos(\phi)-\cos(4\theta+\phi)\\
 & + & 2\sin(2\theta)+\sin(4\theta-\phi)+\sin(4\theta+\phi)+\cos^{5}(\Theta)\sin^{3}(\Theta)(2\cos(2\theta)(-1+\cos(\phi))+\cos(\phi)(-2-2\cos(4\theta)\\
 & + & \cos(4\theta-\phi)-2\cos(\phi)+\cos(4\theta+\phi)-\sin(4\theta)+\sin(4\theta-\phi)+\sin(4\theta+\phi)))\},
\end{array}
\]
\[\begin{array}{lcl}
E & = & \left\{ 1/\left[-8-\cos(2\Theta-\phi)+\cos(6\Theta-\phi)+\cos(4\Theta-2\theta-\phi)-\cos(4\Theta+2\theta-\phi)-\cos(2\Theta+\phi)\right.\right.\\
 & + & \cos(6\Theta+\phi)+\cos(4\Theta-2\theta+\phi)-\cos(4\Theta+2\theta+\phi)+2\sin(4\Theta-2\theta)+2\sin(2(2\Theta+\theta))\\
 & + & \left.\left.2\sin(2\Theta-2\theta-\phi)+2\sin(2\Theta+2\theta-\phi)+2\sin(2\Theta-2\theta+\phi)+2\sin(2\Theta+2\theta+\phi)\right]\right\} \\
 & \times & \{\cos^{8}(\Theta)+\cos^{4}(\theta)\sin^{8}(\Theta)-2\cos^{2}(\theta)\cos(2\phi)\sin^{8}(\Theta)\sin^{2}(\theta)+\sin^{8}(\Theta)\sin^{4}(\theta)\\
 & + & 2\cos^{2}(\Theta)\sin^{6}(\Theta)(-2\cos^{4}(\theta)\cos(\phi)+4\cos^{3}(\theta)\cos^{2}(\phi)\sin(\theta)+\cos^{2}(\theta)(3+\cos(2\phi))\\
 & \times & \sin^{2}(\theta)-4\cos(\theta)\cos(\phi)\sin^{3}(\theta)-2(-1+\cos(\phi))\sin^{4}(\theta))+\frac{3}{32}\sin^{4}(2\Theta)\sin^{2}(2\theta)\\
 & - & 2\cos^{7}(\Theta)\sin(\Theta)(\cos(2\theta)(1+\cos(\phi))+\cos(\phi)\sin(2\theta))-2\cos(\Theta)\cos(2\theta)\sin^{7}(\Theta)\\
 & \times & (-1+\cos(\phi)(1+\sin(2\theta)))+4\cos^{6}(\Theta)\sin^{2}(\Theta)(\cos^{4}(\theta)(1+\cos(\phi))-4\cos(\theta)\cos^{2}(\frac{\phi}{2})\cos(\phi)\sin^{3}(\theta)\\
 & + & \cos(\phi)\sin^{4}(\theta)+\cos^{2}(\theta)\sin^{2}(\theta)\sin^{2}(\phi))-\frac{1}{8}\cos^{4}(\Theta)\sin^{4}(\Theta)(-12-4\cos(4\theta)+\cos(4\theta-2\phi)\\
 & - & 2\cos(2\phi)+\cos(2(2\theta+\phi))+8\sin(4\theta)+4\sin(4\theta-2\phi)+16\sin(2\theta-\phi)+8\sin(4\theta-\phi)+16\sin(2\theta+\phi)\\
 & + & 4\sin(2(2\theta+\phi))+8\sin(4\theta+\phi)+\cos^{3}(\Theta)\sin^{5}(\Theta)(-2\cos(2\theta)(-1+\cos(\phi)+\cos(\phi)(2+2\cos(4\theta)\\
 & + & \cos(4\theta-\phi)-2\cos(\phi)+\cos(4\theta+\phi)+2\sin(2\theta)+\sin(4\theta-\phi)+\sin(4\theta+\phi)))+\cos^{5}(\Theta)\sin^{3}(\Theta)\\
 & \times & (-2\cos(2\theta)(1+\cos(\phi))+\cos(\phi)(-2-2\cos(4\theta)-\cos(4\theta-\phi)+2\cos(\phi)-\cos(4\theta+\phi)+\sin(4\theta)\\
 & + & \sin(4\theta-\phi)+\sin(4\theta+\phi)))\},
\end{array}
\]
\[\begin{array}{lcl}
F & = & \left\{ 1/\left[8-\cos(2\Theta-\phi)+\cos(6\Theta-\phi)+\cos(4\Theta-2\theta-\phi)-\cos(4\Theta+2\theta-\phi)-\cos(2\Theta+\phi)\right.\right.\\
 & + & \cos(6\Theta+\phi)+\cos(4\Theta-2\theta+\phi)-\cos(4\Theta+2\theta+\phi)-2\sin(4\Theta-2\theta)-2\sin(2(2\Theta+\theta))\\
 & + & \left.\left.2\sin(2\Theta-2\theta-\phi)+2\sin(2\Theta+2\theta-\phi)+2\sin(2\Theta-2\theta+\phi)+2\sin(2\Theta+2\theta+\phi)\right]\right\} \\
 & \times & \{\cos^{8}(\Theta)+\cos^{4}(\theta)\sin^{8}(\Theta)-2\cos^{2}(\theta)\cos(2\phi)\sin^{8}(\Theta)\sin^{2}(\theta)+\sin^{8}(\Theta)\sin^{4}(\theta)\\
 & + & 2\cos^{4}(\Theta)\sin^{4}(\Theta)(\cos^{4}(\theta)-2\cos^{3}(\theta)(1-4\cos(\phi)+\cos(2\phi))\sin(\theta)+\cos^{2}(\theta)\cos(2\phi)\sin^{2}(\theta)\\
 & + & 4\cos(\theta)\cos^{2}(\phi)\sin^{3}(\theta)+\sin^{4}(\theta))+2\cos^{2}(\Theta)\sin^{6}(\Theta)(2\cos^{4}(\theta)\cos(\phi)+4\cos^{3}(\theta)\cos^{2}(\phi)\sin(\theta)\\
 & + & \cos^{2}(\theta)(3+\cos(2\phi))\sin^{2}(\theta)+4\cos(\theta)\cos(\phi)\sin^{3}(\theta)+4\cos^{2}(\frac{\phi}{2})\sin^{4}(\theta))+\frac{3}{32}\sin^{4}(2\Theta)\sin^{2}(\theta)\\
 & + & 2\cos^{7}(\Theta)\sin(\Theta)(\cos(2\theta)(-1+\cos(\phi))+\cos(\phi)\sin(2\theta))+2\cos(\Theta)\cos(2\theta)\sin^{7}(\Theta)(1+\cos(\phi)\\
 & \times & (1+\sin(2\theta)))-4\cos^{6}(\Theta)\sin^{2}(\Theta)(\cos^{4}(\theta)(-1+\cos(\phi))+\cos(\phi)\sin^{4}(\theta)-4\cos(\theta)\cos(\phi)\sin^{3}(\theta)\sin^{2}(\frac{\phi}{2})\\
 & - & \cos^{2}(\theta)\sin^{2}(\theta)\sin^{2}(\phi))+\cos^{3}(\Theta)\sin^{5}(\Theta)(2\cos(2\theta)(1+\cos(\phi))+\cos(\phi)(-2-2\cos(4\theta)+\cos(4\theta-\phi)\\
 & - & 2\cos(\phi)+\cos(4\theta+\phi)-2\sin(2\theta)+\sin(4\theta-\phi)+\sin(4\theta+\phi)))+\cos^{5}(\Theta)\sin^{3}(\Theta)(2\cos(2\theta)(-1+\cos(\phi))\\
 & + & \cos(\phi)(2+2\cos(4\theta)-\cos(4\theta-\phi)+2\cos(\phi)-\cos(4\theta+\phi)-\sin(4\theta)+\sin(4\theta-\phi)+\sin(4\theta)+\phi)))\},
\end{array}
\]
\[\begin{array}{lcl}
G & = & \left\{ 1/\left[8-\cos(2\Theta-\phi)+\cos(6\Theta-\phi)-\cos(4\Theta-2\theta-\phi)+\cos(4\Theta+2\theta-\phi)-\cos(2\Theta+\phi)\right.\right.\\
 & + & \cos(6\Theta+\phi)-\cos(4\Theta-2\theta+\phi)+\cos(4\Theta+2\theta+\phi)+2\sin(4\Theta-2\theta)+2\sin(4\Theta+2\theta))\\
 & - & \left.\left.2\sin(2\Theta-2\theta-\phi)-2\sin(2\Theta+2\theta-\phi)-2\sin(2\Theta-2\theta+\phi)-2\sin(2\Theta+2\theta+\phi)\right]\right\} \\
 & \times & \{\cos^{8}(\Theta)+\cos^{4}(\theta)\sin^{8}(\Theta)-2\cos^{2}(\theta)\cos(2\phi)\sin^{8}(\Theta)\sin^{2}(\theta)+\sin^{8}(\Theta)\sin^{4}(\theta)\\
 & + & 2\cos^{4}(\Theta)\sin^{4}(\Theta)(\cos^{4}(\theta)-4\cos^{3}(\theta)\cos^{2}(\phi)\sin(\theta)+\cos^{2}(\theta)\cos(2\phi)\sin^{2}(\theta)\\
 & + & 2\cos(\theta)(1-4\cos(\phi)+\cos(2\phi))\sin^{3}(\theta)+\sin^{4}(\theta))+2\cos^{2}(\Theta)\sin^{6}(\Theta)(2\cos^{4}(\theta)(1+\cos(\phi))\\
 & - & 4\cos^{3}(\theta)\cos(\phi)\sin(\theta)+\cos^{2}(\theta)(3+\cos(2\phi))\sin^{2}(\theta)-4\cos(\theta)\cos^{2}(\phi)\sin^{3}(\theta)+2\cos(\phi)\sin^{4}(\theta))\\
 & + & \cos(\Theta)\cos(2\theta)\sin^{7}(\Theta)(-2+2\cos(\phi)(-1+\sin(2\theta)))+\frac{3}{32}\sin^{4}(2\Theta)\sin^{2}(2\theta)-2\cos^{7}(\Theta)\sin(\Theta)\\
 & \times & (\cos^{2}(\theta)(-1+\cos(\phi))+\cos(\phi)\sin(2\theta))-4\cos^{6}(\Theta)\sin^{2}(\Theta)(\cos^{4}(\theta)\cos(\phi)+(-1+\cos(\phi))\sin^{4}(\theta)\\
 & + & 4\cos^{3}(\theta)\cos(\phi)\sin(\theta)\sin^{2}(\frac{\phi}{2})-\cos^{2}(\theta)\sin^{2}(\theta)\sin^{2}(\phi))+\cos^{3}(\Theta)\sin^{5}(\Theta)(-2\cos(2\theta)(1+\cos(\phi))\\
 & + & \cos(\phi)(-2-2\cos(4\theta)+\cos(4\theta-\phi)-2\cos(\phi)+\cos(4\theta+\phi)+2\sin(2\theta)+\sin(4\theta-\phi)+\sin(4\theta+\phi)))\\
 & + & \cos^{5}(\Theta)\sin^{3}(\Theta)(-2\cos(2\theta)(-1+\cos(\phi))+\cos(\phi)(2+2\cos(4\theta)-\cos(4\theta-\phi)+2\cos(\phi)\\
 & - & \cos(4\theta+\phi)-\sin(4\theta)+\sin(4\theta-\phi)+\sin(4\theta+\phi)))\},
\end{array}
\]
and
\[\begin{array}{lcl}
H & = & \left\{ 1/\left[8+\cos(2\Theta-\phi)-\cos(6\Theta-\phi)+\cos(4\Theta-2\theta-\phi)-\cos(4\Theta+2\theta-\phi)+\cos(2\Theta+\phi)\right.\right.\\
 & - & -\cos(6\Theta+\phi)+\cos(4\Theta-2\theta+\phi)-\cos(4\Theta+2\theta+\phi)+2\sin(4\Theta-2\theta)+2\sin(4\Theta+2\theta)\\
 & + & \left.\left.2\sin(2\Theta-2\theta-\phi)+2\sin(2\Theta+2\theta-\phi)+2\sin(2\Theta-2\theta+\phi)+2\sin(2\Theta+2\theta+\phi)\right]\right\} \\
 & \times & \{\cos^{8}(\Theta)+\cos^{4}(\theta)\sin^{8}(\Theta)-2\cos^{2}(\theta)\cos(2\phi)\sin^{8}(\Theta)\sin^{2}(\theta)+\sin^{8}(\Theta)\sin^{4}(\theta)\\
 & + & 2\cos^{4}(\Theta)\sin^{4}(\Theta)(\cos^{4}(\theta)-4\cos^{3}(\theta)\cos^{2}(\phi)\sin(\theta)+\cos^{2}(\theta)\cos(2\phi)\sin^{2}(\theta)\\
 & + & 2\cos(\theta)(1+4\cos(\phi)+\cos(2\phi))\sin^{3}(\theta)+\sin^{4}(\theta))+2\cos^{2}(\Theta)\sin^{6}(\Theta)(-2\cos^{4}(\theta)(-1+\cos(\phi))\\
 & + & 4\cos^{3}(\theta)\cos(\phi)\sin(\theta)+\cos^{2}(\theta)(3+\cos(2\phi))\sin^{2}(\theta)-4\cos(\theta)\cos^{2}(\phi)\sin^{3}(\theta)-2\cos(\phi)\sin^{4}(\theta))\\
 & - & \cos(\Theta)\cos(2\theta)\sin^{7}(\Theta)(2+2\cos(\phi)(-1+\sin(2\theta)))+\frac{3}{32}\sin^{4}(2\Theta)\sin^{2}(2\theta)+2\cos^{7}(\Theta)\sin(\Theta)\\
 & \times & (\cos^{2}(\theta)(1+\cos(\phi))+\cos(\phi)\sin(2\theta))-4\cos^{6}(\Theta)\sin^{2}(\Theta)(\cos^{4}(\theta)\cos(\phi)+4\cos^{3}(\theta))\cos^{2}(\frac{\phi}{2})\\
 & \times & \cos(\phi)\sin(\theta)+2\cos^{2}(\frac{\phi}{2})\sin^{4}(\theta)+\cos^{2}(\theta)\sin^{2}(\theta)\sin^{2}(\phi))+\cos^{3}(\Theta)\sin^{5}(\Theta)(2\cos(2\theta)(-1+\cos(\phi))\\
 & + & \cos(\phi)(2+2\cos(4\theta)+\cos(4\theta-\phi)-2\cos(\phi)+\cos(4\theta+\phi)-2\sin(2\theta)+\sin(4\theta-\phi)+\sin(4\theta+\phi)))+\\
 & + & \cos^{5}(\Theta)\sin^{3}(\Theta)(2\cos(2\theta)(1+\cos(\phi))+\cos(\phi)(-2-2\cos(4\theta)-\cos(4\theta-\phi)+2\cos(\phi)-\cos(4\theta+\phi)\\
 & + & \sin(4\theta)+\sin(4\theta-\phi)+\sin(4\theta+\phi)))\}.
\end{array}\]

\end{document}